\documentclass[%
 reprint, 
 superscriptaddress,
nofootinbib,
nobibnotes,
 amsmath,amssymb,
 aps,
 pra,
floatfix,
]{revtex4-2}

\usepackage{graphicx}
\usepackage{dcolumn}
\usepackage{multirow}
\usepackage[table]{xcolor} 
\usepackage{bm}
\usepackage{orcidlink}
\usepackage{hyperref}
\usepackage{colortbl}
\usepackage{makecell}

\usepackage[textsize=footnotesize,backgroundcolor=red!25]{todonotes}

\definecolor{SetA}{HTML}{1F77B4}
\definecolor{SetB}{HTML}{FF7F0E}
\definecolor{SetC}{HTML}{2CA02C}
\definecolor{SetD}{HTML}{D62728}
\definecolor{SetE}{HTML}{9467BD}
\definecolor{SetF}{HTML}{8C500B}
\definecolor{SetG}{HTML}{E377C2}
\definecolor{SetH}{HTML}{17BECF}
\definecolor{SetA1}{HTML}{06D6A0} 
\definecolor{SetB1}{HTML}{4A2C2A} 
\definecolor{SetC1}{HTML}{FFBE0B} 
\definecolor{SetD1}{HTML}{1111EC} 

\begin{document}

\title{Using domain walls to benchmark fast quantum anneals}

\author{Shushmi Chowdhury\orcidlink{0009-0005-7658-713X}}
\email{shushmi.chowdhury@strath.ac.uk}

\author{Asa Hopkins}
\email{asa.hopkins@strath.ac.uk}
\affiliation{Department of Physics, University of Strathclyde, Glasgow, G4 0NG, United Kingdom.}

\author{Nicholas Chancellor\orcidlink{0000-0002-1293-0761}}
\email{nick.chancellor@newcastle.ac.uk}
\affiliation{School of Computing, Newcastle University, Newcastle upon Tyne NE1 7RU, United Kingdom.}

\author{Viv Kendon\orcidlink{0000-0002-6551-3056}}
\email{viv.kendon@strath.ac.uk}
\affiliation{Department of Physics, University of Strathclyde, Glasgow, G4 0NG, United Kingdom.}

\begin{abstract}
We consider the distribution of domain walls arising in ground states of frustrated 1-dimensional Ising spin chains with anti-parallel boundary spins. For ideal chains, a sine-squared \emph{particle-in-a-box} distribution of domain walls is expected in the quantum setting, and a uniform distribution expected classically.  However, for chains with random imperfections, prior work [Chancellor et al., npj Quantum Inf.~8, 73, 2022] found that the distribution in a physical implementation corresponded to neither. Instead, the domain wall distribution forms a U-shaped profile arising from a classical model with random fields, where the combination of averaging over these fields and thermal averaging produces the characteristic U-shape. With D-Wave's newer fast anneal feature, we access anneal times shorter than the coherence time and observe the expected coherent quantum behavior in the form of a sine-squared distribution. For the shortest available anneal time, the dynamics have not yet reached the sine-squared shape and instead display a different U-shape in the middle of the chains.  This motivates using the extent of the domain wall distribution ``tail'' in this regime as a platform independent metric that provides a signature of both coherence time and noise levels. 
\end{abstract}

\maketitle

\section{Introduction}
\label{sec:intro}
Quantum computing promises greater efficiency for solving many types of industrially relevant problems.  Current quantum hardware is still at a relatively early stage, with multiple types of general and special purpose quantum devices in active development.  General purpose quantum computers are usually based on the circuit model, where discrete quantum gates are applied to a register of qubits (two state quantum systems).  The circuit model of quantum computing is \emph{universal}, capable of efficiently simulating other models of universal quantum computing, such as quantum walks \cite{childs09a,childs12a} or measurement based quantum computing \cite{raussendorf01a}.  Special purpose quantum devices are also under development, notably for quantum simulation \cite{PRXQuantum.2.017003}, 
and for solving classical optimization problems.  These are designed to be efficient for a particular class of problems, trading generality for efficiency in their chosen domain.

Many combinatorial optimization problems can be mapped onto a search for the lowest energy configuration of a network of discrete spins interacting via an Ising Hamiltonian \cite{FINNILA1994343,PhysRevE.58.5355}. 
Special purpose quantum hardware has been built to do this \cite{DWave_schedule_docs,Levine_2019} 
using quantum and thermal fluctuations to increases the likelihood of locating low energy configurations in complex landscapes with many local minima.  Such quantum annealers can outperform other approaches both in success rate and in residual energy scaling \cite{Santoro_2002}, thus suggesting a scalable route to solving hard optimization problems \cite{QAmat25, 10.1038/nature10012, PhysRevLett.134.160601}.

This has motivated exploration of other approaches that drive the quantum Ising systems faster.  Diabatic quantum annealing (DQA) trades strict ground-state evolution for coherent dynamics through excited states, currently is not known to be efficiently simulable by classical algorithms and in some specific problems, such as MAX-2-SAT, has been shown to outperform adiabatic QA \cite{2021NatRP...3..466C}. 
DQA can serve as a powerful tool to simulate complex non-equilibrium dynamics \cite{article23} at a fundamental level. With state-of-the-art flux-qubit devices and emerging neutral atom platforms now able to access diabatic regimes \cite{Kendon_2026}, researchers can study phase transitions, magnetic frustration, and many-body phenomena at scales difficult for classical computers to emulate.

In a recent benchmarking review led by the UK National Physical Laboratory (NPL) \cite{lall2025reviewcollectionmetricsbenchmarks}, a single qubit control error metric for quantum annealing was documented, an extension of an approach introduced earlier in \cite{Chancellor_2022}. The metric uses short spin chains with pinned ends to enforce frustration, and estimates the noise strength of random longitudinal fields present in the annealer hardware by fitting experimentally obtained domain wall position distributions to a numerically computed Boltzmann distribution averaged over many noise realizations. 
The combined effect of averaging over these thermal fluctuations arising from finite, non-zero temperature of the system, and over many random noise realizations produces a characteristic U-shaped distribution. 

Other benchmarks using domain walls in spin chains have been proposed, e.g., \citeauthor{pelofske2025erasingclassicalmemoryquantum} \cite{pelofske2025erasingclassicalmemoryquantum} consider the movement of such domain walls and the entropy generated.  
Work carried out by \citeauthor{Chancellor_2019} \cite{Chancellor_2019} demonstrated that the position of a single domain wall in a ferromagnetic Ising chain can be used to encode discrete variables for use in optimization problems. 
Benchmarks based on spin chains provide a simple, hardware independent metric for quantum annealers that can be implemented on any hardware that can instantiate an Ising spin chain, such as neutral atoms \cite{Levine_2019}.   Other specliazed hardware for solving optimization problems includes photonic devices \cite{slysz2024}.  

General purpose quantum computers can simulate quantum annealing to solve optimization problems and this has been extensively studied as the quantum approximate optimization algorithm (QAOA) or the quantum alternating operator ansatz \cite{Farhi2014a,Farhi2014b,Marwaha2022}.  Implementing spin chain benchmarks in a QAOA setting provides for cross-platform comparisions between special and general purpose hardware.  The trade off between imperfect resources on special and general purpose hardware for different applications is especially important to assess in the NISQ and early fault tolerant eras.

For our work in this paper, we used the most mature available hardware in 2025 for quantum annealing, specifically, a D-Wave Advantage 4.1 QPU.  
In \cite{Chancellor_2022}, the spin chain benchmark was experimentally implemented on the Vesuvius 6 D-Wave QPU (quantum processing unit), built on the \emph{Chimera} topology. Here, we reproduce the characteristic U-shaped distribution by implementing the same problem Hamiltonian on the newer D-Wave Advantage 4.1 QPU, which employs the \emph{Pegasus} topology with enhanced qubit connectivity, and provides shorter anneal times than previous hardware generations. Moreover, and as initially expected but not observed in \cite{Chancellor_2022}, we do observe a particle-in-a-box distribution for suitable short anneal times. This is characterized by a sine-squared profile, and was achieved by making use of the \emph{fast anneal} protocol of the newer D-Wave hardware \cite{King:2022phl,article23}.

These observations motivate using the tail of the domain wall distribution (the part decreasing towards the ends of the chains) as a probe of coherence and noise levels in analogue quantum devices that can instantiate Ising model Hamiltonians. In the coherent regime, the domain wall behaves as an effective quantum particle whose wavefunction spreads coherently along the chain. Annealing very fast will disrupt the interference by not providing enough time for the equilibrium distribution to be reached.  Annealing more slowly will increase the impacts from noise and thermal effects.  With long enough coherence and low enough noise, it will be possible to find an anneal time long enough to reach equilibrium, while still remaining minimally affected by noise levels, to produce the characteristic sine-squared shape. Since the sine-squared shape is expected to show the longest possible tails, we propose the tail length as a simple and robust metric for assessing performance of analogue quantum hardware. 

In this work, we present a detailed study on D-Wave Systems that supports this proposed metric. In section \ref{sec:theory}, we provide the background for the spin chain behavior in a quantum annealer. In section \ref{sec:methods}, we explain our methodology for analyzing the experiments and comparing with numerical simulations. Section \ref{sec:expt} presents the results from our experiments and comparisons with numerical simulations. In section \ref{sec:tail metric}, we propose a metric based on the tail of the obtained probability distribution, and provide further simulations to characterize its properties. We discuss and conclude in \ref{sec:discuss} and provide technical details in a series of appendices.

\section{Theoretical models}
\label{sec:theory}

The general time dependent Hamiltonian representing the transverse field Ising model \cite{Pfeuty1970}, in the form implemented by D-wave hardware, is given by
\begin{equation}\label{eq:HTI}
    H(t)=-\frac{A(s)}{2}\sum_{i=1}^{L}\sigma_i^{x} + \frac{B(s)}{2}H_{\text{problem}},
\end{equation}
where $\sigma_i^{x}$ is the Pauli-x operator applied to the $i$th spin in a total of $L$ spins, $H_{\text{problem}}$ is the Ising spin Hamiltonian encoding the problem to be solved, and $A(s)$, $B(s)$ are real-valued, time dependent annealing schedules. D-Wave documents the energy scales of $A(s)$ and $B(s)$ using frequency in GHz, in units where Planck's constant is equal to one. The dimensionless annealing parameter $s=t/t_f$, where $t$ is
the annealing time in microseconds, and $t_f$ is the total duration of the anneal. 
The system is initialized in the ground state of the transverse field $\sum_{i=1}^{L}\sigma_i^{x}$, i.e., all spins lined up in the $x$ direction. Over time, $A(s)$ is decreased and $B(s)$ is increased, gradually turning off the dynamics driven by the transverse field and turning on the problem Hamiltonian, so the system ideally ends in the ground state of $H_{\text{prob}}$. 
\begin{figure}[tb!]
\includegraphics[width=1\linewidth]{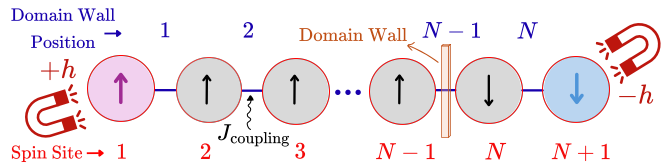}
\caption{\label{fig:spin_chain}  Schematic showing an Ising chain of $L=N+1$ spins with twisted boundary conditions imposed by longitudinal fields $\pm h$ on spins $0$ and $N+1$, with a single domain wall between spin sites $N-1$ and $N$. The $N$ possible domain wall positions are labeled above and the $N+1$ spins are labeled below. Ferromagnetic coupling strengths $J$ are applied between neighboring spins.}
\end{figure}

\subsection{Frustrated spin chains}
\label{ssec:fchains}
A particularly simple, but insightful, example of a physical spin system is a finite ferromagnetic spin chain in which the spins at the two ends of the chain are constrained to have opposite polarizations. This introduces an inherent frustration into the system, ensuring that at least one domain wall between opposite spin polarizations must be present in the chain. 
The problem Hamiltonian representing a one dimensional ferromagnetic Ising chain of length $L=N+1$ with anti-parallel boundary spins (see Fig.~\ref{fig:spin_chain}) is given by 
\begin{equation}\label{eq:Hspinchain}
    H_{\text{chain}} = J\sum_{i=1}^{N}-\sigma_i^z\sigma_{i+1}^z + h(\sigma_1^z-\sigma_{N+1}^z),
\end{equation}
where $\sigma_i^{z}$ is the Pauli-z operator applied to the $i$th spin, $J$ is the coupling strength between neighboring spins, and $h$ is the strength of the longitudinal fields pinning spins $1$ and $N+1$ in opposite polarizations. 

\begin{figure}[tb!]
\includegraphics[width=0.95\linewidth]{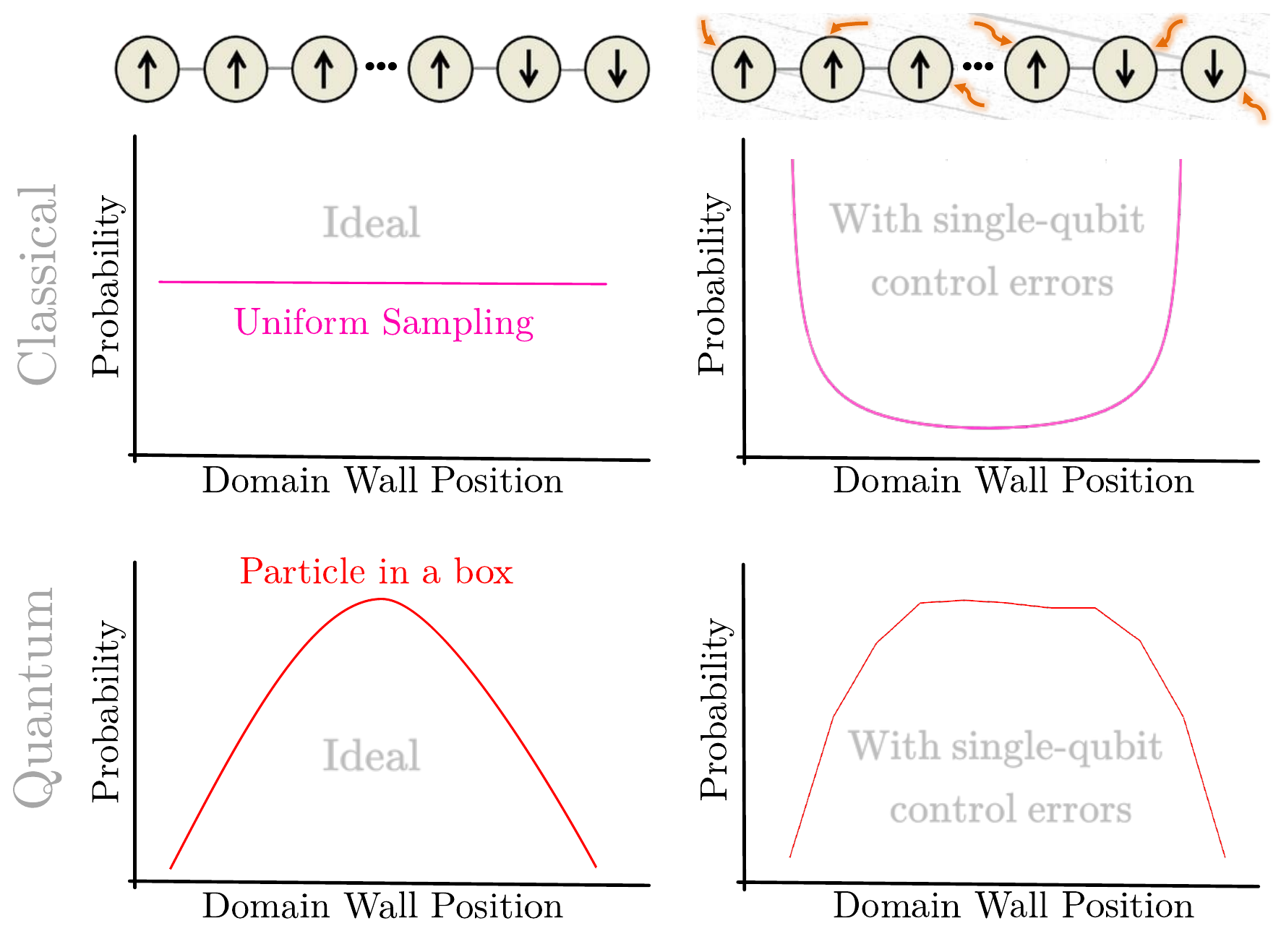}
\caption{\label{fig:field_errors} 
Illustration of the expected domain wall distributions for classical (top row) and quantum (bottom row) ideal (left) and with random fields (right) frustrated spin chains in their ground states (single domain wall).
}
\end{figure}

In the quantum setting where a transverse field is applied, as in Eq.~\eqref{eq:HTI}, the spins are driven to rotate between up and down states. As a result, a single domain wall behaves like a delocalized quantum particle obeying the particle-in-a-box model, as illustrated in Fig.~\ref{fig:field_errors} bottom left. The ground state follows a sine-squared distribution, with the highest probability near the middle of the chain and vanishing at the ends. 
In the classical limit, the domain wall can reside with equal probability on any of the $N$ bonds in an $N+1$ spin chain system. This is because all bonds are energetically equivalent (degenerate), giving no preferred location to the domain wall, Fig.~\ref{fig:field_errors} top left. 
During the quantum annealing process, domain walls appear as the system undergoes a quantum phase transition from a quantum paramagnetic phase to an ordered ferromagnetic phase at the critical point, where $A(s_c) = B(s_c)$ in Eq.~\eqref{eq:HTI}. 

\subsection{Residual random fields}
\label{ssec:res_fields}
\citeauthor{Chancellor_2022} \cite{Chancellor_2022} found that neither of the ideal distributions shown in Fig.~\ref{fig:field_errors} left column were obtained using D-Wave hardware. Instead, programming the time dependent Hamiltonian given in Eq.~\eqref{eq:HTI} with the frustrated chain Hamiltonian from Eq.~\ref{eq:Hspinchain} onto the D-Wave annealer and averaging over a large number of runs produced a U-shaped profile like Fig.~\ref{fig:field_errors} top right.  As shown theoretically and numerically in \cite{Chancellor_2022}, this shape results from a combination of thermal averaging and disorder averaging. The former stems from stochastic fluctuations induced by the hardware's coupling to a thermal bath.  The latter is attributed to the residual random longitudinal fields in the annealer, originating from physical hardware imperfections, that cannot be removed by the callibration processes.

To model these residual fields, an additional term of the form 
\begin{equation}\label{eq:ran_fields}
    H_{\mathrm{fields}} = \sum_{i=1}^{L=N+1} \zeta_i \sigma_i^z
\end{equation}
is added to the total Hamiltonian in Eq.~\eqref{eq:HTI}.
The $\zeta_i$ are uncorrelated random variables that are Gaussian distributed 
$\zeta_i \in \mathcal{N}(0,\bm\zeta^2)$
with zero mean and variance $\bm{\zeta}^2$, and they are much weaker than the coupling strength between spins, i.e., $\bm\zeta \ll B(s_c)J$ where $\bm\zeta$ is also in energy units in GHz.
The random fields $\zeta_i$ are assumed to be constant over the time scale of the system dynamics, and act locally on the $i$th qubit in the Ising spin chain. The total time-dependent Hamiltonian is thus
\begin{equation}
H(t)=-\frac{A(s)}{2}\sum_{i=1}^N\sigma_i^{x}  + \frac{B(s)}{2}H_{\text{chain}} + H_{\text{fields}}. 
\label{eq:H_res_fields}
\end{equation} 

Adding these random longitudinal fields lifts the degeneracy of the degenerate energy levels of the ideal Hamiltonian. Suppose the domain wall is at position $N-1$, then moving the domain wall to position $N$ flips the orientation of spin $N$ (refer to Fig.~\ref{fig:spin_chain}).  For a given noise realization, the random field energy associated with spin $N$ changes from $-\zeta_{N}$ to $+\zeta_{N}$, thus changing the energy of $H_{\text{fields}}$ by $2 \zeta_{N}$.  For any arbitrary domain wall positions $n>m$, where m and n label the domain wall positions and $i$ labels the spin sites, 
moving the domain wall from position $m$ to $n$, therefore changes the state energy by
\begin{equation}
E_n - E_m = 2\!\sum_{i = m+1}^{n}\zeta_i.
\label{eq:move_domain_wall}
\end{equation}

\subsection{Thermalized random fields}
In a thermalized regime with a single domain wall present, the occupation of domain wall positions will follow a Boltzmann distribution with the probability of a domain wall at position $n$ given by 
\begin{equation}\label{eq:PnE}
    P_n =\frac{\exp{\left\{-\beta E_n \right\}}}{\sum_{i=1}^N \exp{\left\{-\beta E_i \right\}}} 
\end{equation} 
where $\beta$ is the inverse temperature and $E_n$ is the energy of the chain with a single domain wall in position $n$. 
Using Eq.~\eqref{eq:move_domain_wall}, the energies of each domain wall location can be written as
\begin{equation}\label{eq:E}
    E_n = E_1 + 2\sum_{j = 2}^n \zeta_j,
\end{equation}
which in the thermal regime gives a probability distribution of
\begin{equation}\label{eq:Pnzeta}
    P_n =\frac{1}{Z}\exp{\left\{-\beta E_1 - 2\beta\sum_{j=2}^n\zeta_j \right\}}
\end{equation} 
where $Z=\sum_i\exp(-\beta E_i)$ is the partition function.

We now need to average this Boltzmann distribution over the random distribution of $\zeta_j$ for disorder averaging.  We do this numerically using Monte Carlo simulation to estimate the average of Eq.~\eqref{eq:Pnzeta}, further details of our numerical methods are given in section \ref{ssec:numsim}. The resulting average distribution is U-shaped for low temperatures (large $\beta$), as illustrated in Fig.~\ref{fig:field_errors} (top right). 

In the limit as the temperature goes to zero, the distribution approaches the arcsine distribution, as can be seen as follows:
Eq.~\eqref{eq:E} implies that the energy levels follow a random walk: each energy level is equal to the previous one plus a random step. It is known from random walk statistics that the location of the energy minimum is distributed according to an arcsine distribution \cite{arcsine}. As $\beta \rightarrow \infty$, this distribution is recovered by the ensemble average over Eq.~\eqref{eq:PnE}. It is also the expected distribution in a fully coherent spin chain as the evolution time approaches infinity, according to the adiabatic theorem. At high temperatures ($\beta \rightarrow 0$) the distribution flattens into the classical distribution shown top left, as expected when the random fields are negligible compared to the thermal fluctuations.

\subsection{Diabatic regime}
\label{ssec:coherent}
Due to our assumption of a Boltzmann distribution in Eq.~\eqref{eq:PnE}, the averaged distribution obtained by averaging Eq.~\eqref{eq:Pnzeta} is for a fully thermalized chain with a single domain wall, and is not valid in the diabatic regime, i.e., at short times, during coherent evolution or when a mixture of coherent and thermal effects are present.  The accessible coherent regime in D-Wave hardware has been well studied recently \cite{King:2022phl,article23}, with the crossover to thermalized behavior occuring between 20 ns and 30 ns in the D-Wave hardware we used \cite{article23}.  Our Monte Carlo simulations suggest that the adiabatic limit is approached at around 20 ns, so for more coherent hardware, such as neutral atoms, both diabatic and adiabatic regimes should be accessible \cite{Kendon_2026}.

For short chains, such as those we consider in this work, the dynamics of domain wall formation is well captured by a single Landau–Zener transition \cite{Zener:1932ws}, a two-level model describing the probability of excitation when an avoided energy-level crossing is traversed at a finite rate \cite{Ivakhnenko2022}. 
For the fastest anneal times available on the D-Wave systems we used, a significant fraction of the runs end with more than one domain wall present (see table~\ref{tab:domain_count} in appendix \ref{app:data}).  We use numerical methods to characterise this regime, described in more detail in section \ref{ssec:numsim}.
By evolving the Hamiltonian in Eq.~\eqref{eq:H_res_fields} under the time dependent Schr\"odinger equation following D-Wave's annealing schedule \cite{DWave_schedule_docs}, we can predict the expected number of multiple domain wall outcomes to compare with our experimental data.

\section{Methods} 
\label{sec:methods} 

To test the theoretical predictions for fast, coherent anneals, we performed experiments using newer D-Wave system hardware than was available for \cite{Chancellor_2022}, and carried out numerical simulations to fit the experimental data and validate our models of the system behavior.  The choice of short spin chains ensures that the fits and simulations of the models are relatively straightforward and were performed on laptops and desktop workstations.  However, the D-Wave experiments require a number of technical details to be specified and assumptions to be tested.  

\subsection{D-Wave experiments}
\label{ssec:dwave_method}
We implemented the Ising chain Hamiltonian Eq.~\eqref{eq:Hspinchain} on the D-Wave Advantage 4.1 system. A key feature of this hardware, compared to previous systems, is its \emph{fast anneal} protocol, which enables annealing times as short as $0.005\, \mu$s, a significant enhancement over the previous minimum of $0.5\, \mu$s in the \emph{standard} protocol.
All QPU experiments were implemented in Python \cite{Python} using the D-Wave Ocean SDK, with the Ising Hamiltonian constructed as a binary quadratic model (BQM) using the Dimod package, and submitted to an Advantage system 4.1 processor. The hardware connectivity graph was processed and represented using the D-wave extension dwave-networkx, and the python package NetworkX \cite{NetworkX}. All of our D-Wave experiments were carried out during May 2025.  
Experimental data will be made available on \cite{OurRawData}, and summaries of the raw data are given in the tables in appendix \ref{app:data}.

For a standard anneal, a time dependent external flux is applied to create the $h_i \sigma^z_i$ terms in the Hamiltonian in Eq.~\eqref{eq:Hspinchain}, and the time dependent protocol ensures that the strengths of the $h$ and $J$ terms stay in the same proportions throughout the anneal.
In our experiments and simulations, we have set $h=2J$, large enough to ensure the pinned spins stay pinned for most of the spin chains.  Nonetheless, the spins do occasionally flip, and details of these error counts are given in table ~\ref{tab:flip} in appendix \ref{app:data}.

In order to achieve fast anneal times, some changes are made at the hardware level \cite{dwavequantumSolverParameters}.  In the fast anneal protocol the external flux is no longer time dependent, instead being held constant. The constant can be chosen by the user in the form of flux biases \cite{DWave_schedule_docs}.
For our purposes, it is only necessary that $h$ is large enough to enforce the single domain wall as a ground state. We can therefore apply a constant value for the external flux on the first and last qubit corresponding to $h$ in Fig.~\ref{fig:spin_chain}. This is done in the D-Wave software by providing a flux-bias offset for each qubit, along with the necessary unit conversions \cite{DWave_solver_properties_specific}. 
We used the default value for the flux bias, which corresponds to making the fast schedule match the standard anneal schedule at the point $s_c$ where $B(s_c) = A(s_c) = 1.391$ GHz for this particular D-Wave system. Note that the documentation mistakenly writes $s = 0.612$ when this point actually corresponds to $s \approx 0.364$ \cite{DWave_flux_docs}.

\begin{figure}[bt!]
\includegraphics[width=0.85\linewidth]{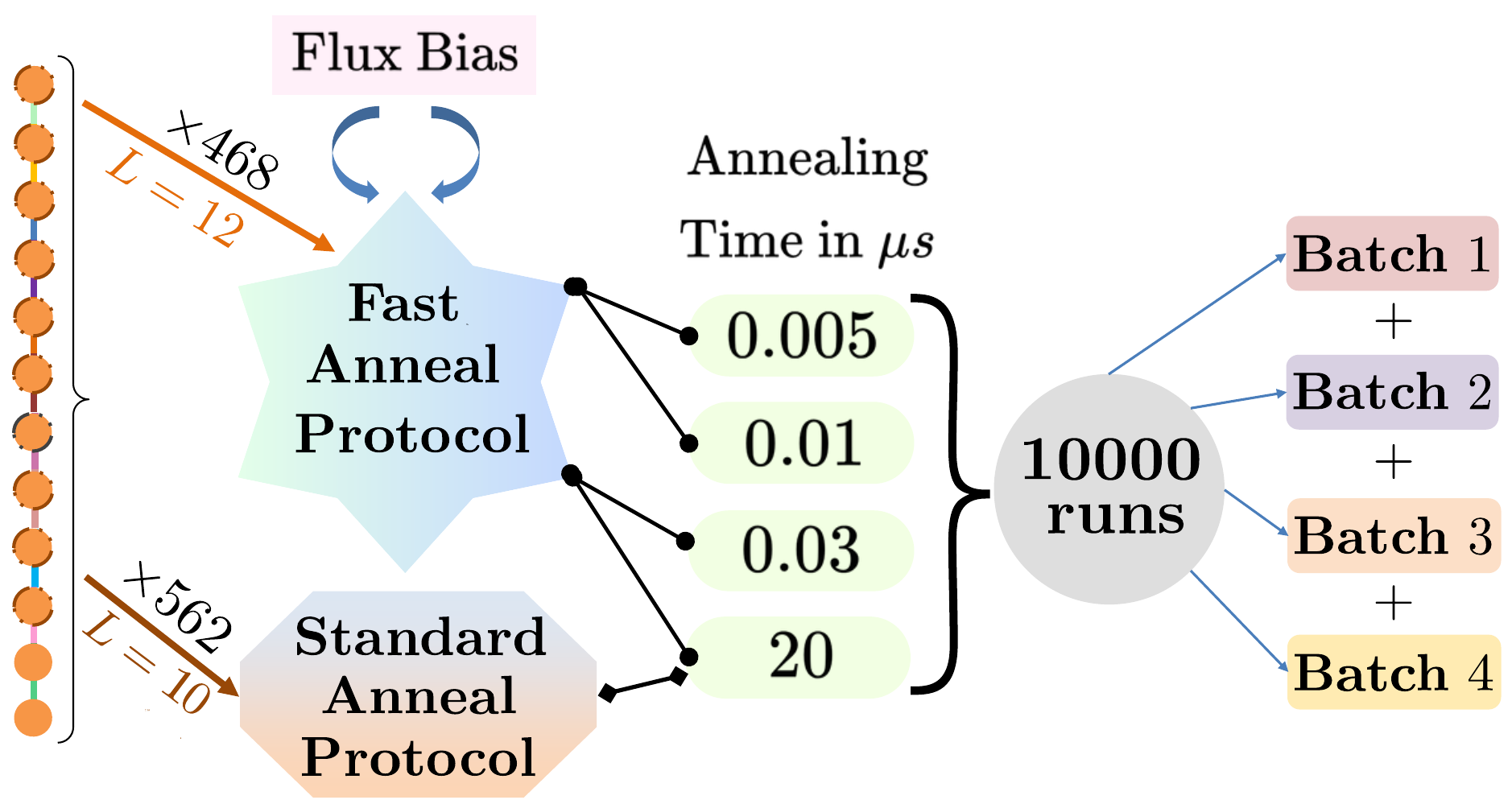}
\caption{\label{fig:figure_outline_methods}  Outline of the experimental settings on the D-Wave quantum annealer. Problem instances are submitted using both the standard anneal protocol with an anneal time of $20\,\mu$s, and the fast anneal protocol with times of $0.005\,\mu$s, $0.01\,\mu$s, $0.03\,\mu$s, and $20\,\mu$s. In the fast anneal protocol, a flux bias is applied since non-zero local fields ($h_i$) cannot be directly specified. A chain length of $L = 10$ was used in the standard protocol with 562 chains embedded across the available hardware qubits, and $L = 12$ in the fast anneal protocol with 468 chains embedded. Due to maximum QPU run time constraint per submission, the $10\,000$ were divided into four batches.
}
\end{figure}

Since the fast anneal and standard protocols work differently, we tested both with the longer anneal time of $20\,\mu$s, to check that they perform equivalently for the spins chains we are studying. 
In the standard protocol, we set the chain length to $L=(N+1)=10$ to allow direct comparison with \cite{Chancellor_2022}, where chain lengths of $L=10$ were used. For the fast anneal protocol, we use a slightly longer chain, $L=12$, which remains small enough for meaningful comparison, but provides slightly better statistics for fitting and interpreting the results. 
We used anneal durations of $0.005\,\mu$s, $0.01\,\mu$s, $0.03\,\mu$s, as well as $20\,\mu$s, using the fast anneal protocol, in order to characterize the diabatic regime, which was not accessible when the work in \cite{Chancellor_2022} was carried out. Unless otherwise noted, each problem instance was run $10^4$ times in the experiments, in four separate batches of between $2000$ and $3000$ each. An overview of the experimental parameters and annealing settings is provided in Fig.~\ref{fig:figure_outline_methods}. 

To make efficient use of the full size of the D-Wave chips, we performed a one-to-one minor embedding by using a heuristic algorithm for finding Hamiltonian paths to map logical to physical qubits, see appendix~\ref{app:emb} for details.   This allowed us to embed 562 chains of length $L=10$ or 468 chains of length $L=12$ 
into a single run.

Systematic effects in D-Wave hardware show up as gauge dependent errors.  In appendix \ref{app:gauge} we show the effects of gauge dependent errors and how they can be removed by \emph{gauge averaging}.  This is a standard error mitigation procedure, and we use it for all runs unless noted in the caption and text.

In appendix \ref{app:single_error}, we show that the standard and fast anneal protocols produce the same U-shaped profile for the domain wall distribution for the same anneal times of 20 $\mu$s.  This confirms that the different mechanisms in the fast anneal protocol are equivalent to the standard protocol in the same regime.

Although the ground state of frustrated spin chains is a state with a single domain wall, the short anneal times do not always reach a ground state configuration.  We therefore post-selected the raw results from D-Wave runs for outcomes with a single domain wall.  In most cases, this only slightly reduced the total number of samples, but for longer chains and very short run times, the reduction is significant. Appendix \ref{app:post_select} contains Fig.~\ref{fig:domain_multi_wall} 
showing the probability distributions with all domain walls and post selected single domain walls. Appendix \ref{app:data} contains tables \ref{tab:domain_count} and \ref{tab:domain_count1} summarizing the post selection details.  The number of data points used after post selection, averaged over all the chains in a single run (four batches) is mentioned in the caption of all figures containing experimental data.
 
\subsection{Numerical simulation methods}\label{ssec:numsim}

The theoretical models from section \ref{sec:theory} are in general not analytically tractable and require numerical simulations to compare with experimental data.  Our simulations were carried out using Python 3.14.4 and C++ compiled with GCC 15.2.1. The main numerical work in Python was done using NumPy 2.4.4 \cite{harris2020array} and SciPy 1.17.1 \cite{2020SciPy-NMeth}, and quantum time evolution was implemented using QuTiP 5.2.3 \cite{qutip5}. Plots are generated using the Python package Matplotlib 3.10.9 \cite{matplotlib}. In C++, calculations were done using Eigen 5.0.1-2 and Vector Class Library 2.02.01. For the sizes of our experiments, a desktop workstation was sufficient for all the simulations. Our code will be made available alongside the experimental and simulation data \cite{OurRawData}. 

The experimental data was fitted using Monte Carlo simulations to calculate the average probability distributions from the theoretical models.  For the strength of the random fields $\bm{\zeta}$, the simulated distribution was calculated as follows: An ensemble of $5\times 10^5$ simulations was calculated over a range of $\bm{\zeta}$ values. For each simulation, the random fields $\zeta_i$ were each sampled from $\mathcal{N}(0,\bm\zeta^2)$, and then the system was evolved under the Hamiltonian in Eq.~\eqref{eq:H_flux_biases} for $0.005\,\mu$s, starting from the uniform superposition. The mean and standard deviation of the probability of each domain wall position was then calculated from this ensemble. To allow our fitting routine to search values between sample points, polynomial interpolation was used, with a polynomial being fit to each domain wall probability individually. To minimise error, the sample points were chosen as the 16 Chebyshev nodes between $[0,\frac{1}{\pi}]$, where the units are GHz. By using the same random seed at each value of $\bm{\zeta}$, this function is smooth and the polynomial fit converges. Numerical error is dominated by the standard error on the ensemble mean, representing a relative error of roughly $0.1\%$. 

To do the fitting, the Kullback–Leibler (KL) divergence between the distribution obtained from the $0.005\,\mu$s experimental data and the ensemble average obtained from the simulations was minimised with respect to $\bm{\zeta}$, which is asymptotically equivalent to maximum-likelihood estimation. To estimate the error on the fit, the experimental data was resampled $10^5$ times and the minimisation repeated. 

The KL divergence is the excess surprisal when using the second distribution to approximate the first, measured in nats per sample (when using $\ln$ rather than $\log_2$). 
This is chosen to make comparisons with \cite{Chancellor_2022,lall2025reviewcollectionmetricsbenchmarks}.

We note a potential limitation of the KL divergence arising from is its sensitivity in the presence of zero probability bins, since the divergence can tend to infinity when the second distribution has a zero probability to an outcome with non zero probability in the first distribution, and indeed this is pointed out in \cite{lall2025reviewcollectionmetricsbenchmarks}. In our study, we did not encounter zeros in the distributions considered. 

To estimate $\beta$, the inverse operating temperature, the same method was used with details as follows:
An ensemble of $10^7$ samples was used, for each sample, the random fields $\zeta_i$ were sampled from $\mathcal{N}(0,\bm{\zeta})$, using the value of $\bm\zeta$ obtained from fitting the fast anneal data, and the domain wall probabilities were calculated from Eq.~\eqref{eq:Pnzeta}. The mean and standard deviation of the probability of each domain wall position is then calculated from this ensemble. A polynomial was fit to each domain wall probability, using $\beta$ at the 16 Chebyshev nodes between $[0,1]$. 
Numerical error is dominated by the standard error on the ensemble mean, representing a relative error of roughly $0.01\%$. Then, the KL divergence between the experimental data and the ensemble average was minimised with respect to $\beta$, with the errors estimated by resampling in the same way as for the random fields.

\section{Experimental Results and Fitting}
\label{sec:expt}

We first characterize the behavior of the short spin chains on D-Wave systems under short anneal times, now that the fast anneal protocol is available.
The domain wall behavior on frustrated spin chains in the thermalized regime (anneal times of $20\,\mu$s) is well understood from the work of \cite{Chancellor_2022}, and we show in appendix \ref{app:single_error} that we can reproduce these results on the newer D-Wave hardware, including for the newer fast anneal protocol run for the same slower anneal times.

\subsection{Domain wall behavior under coherent dynamics}
\label{ssec:fast_anneal}
\begin{figure}[tb!]
\includegraphics[width=0.95\linewidth]{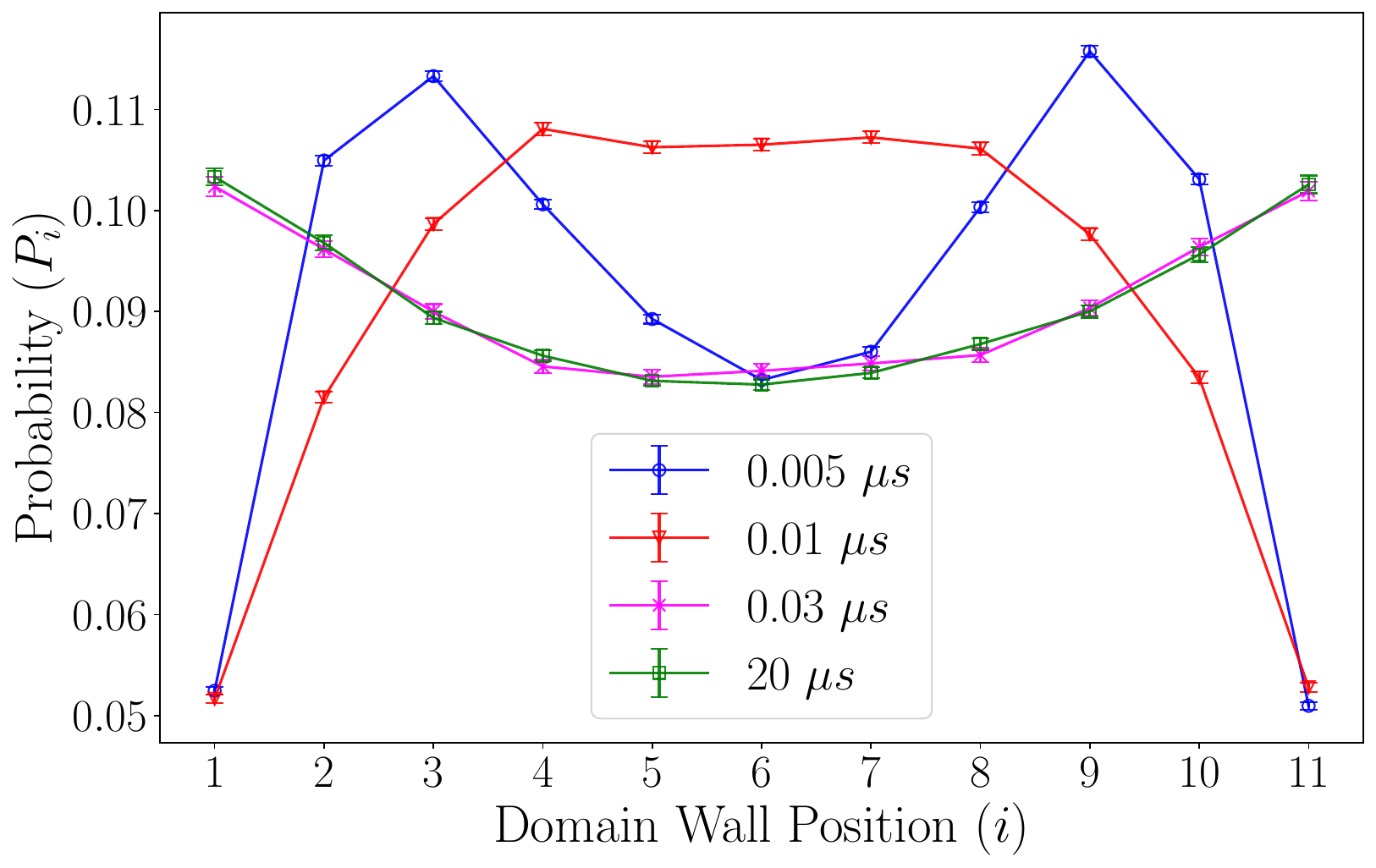}
\caption{\label{fig:fast_anneal} Probability distribution from the fast anneal protocol averaged over 468 chains for a chain length of 12 with $10^4$ runs for different anneal times. The data have been post selected to only show single domain wall instances. On average across 468 chains per run the single  domain wall count is $7072.9 \pm 11.2$ for $0.005\,\mu$s, $9119.4 \pm 17.4$ for $0.01\,\mu$s, $9960.9 \pm 26.7$ for $0.03\,\mu$s, and $9762.9 \pm 21.3$ for $20\,\mu$s (see table~\ref{tab:domain_count}). Error bars represent the uncertainty obtained by combining in quadrature the propagated binomial uncertainty from finite domain wall occurrences and the standard error of the mean across chains.}
\end{figure}

Figure \ref{fig:fast_anneal} shows the domain wall distributions for anneal times of $0.005\,\mu$s, $0.01\,\mu$s, $0.03\,\mu$s and $20\,\mu$s.  We observe a fair approximation of the sine-squared particle-in-a-box profile at $0.01\,\mu$s. This behavior was anticipated in \cite{Chancellor_2022}, but short enough anneal times were not available then. 
The profiles at longer times of $0.03\,\mu$s and $20\,\mu$s show the emergence of the U-shape after thermalization that was reported in \cite{Chancellor_2022}.  The fastest anneal runs, $0.005\,\mu$s yield a slightly skewed M-shaped profile, U-shaped in the middle with turned down ends.  This is quite distinct from the U-shape produced after thermalization in the $0.03\,\mu$s and $20\,\mu$s profiles.  It is more consistent with the domain wall not having sufficient time to localize across the whole chain, and the high number of samples with multiple domain walls ($2,927.1 \pm 11.2$ out of $10^4$ samples, averaged over $468$ chains per run, see table \ref{tab:domain_count} in appendix \ref{app:data}) supports this hypothesis.

\subsection{Parameter fitting}\label{ssec:fits}

To investigate the factors governing the domain wall distribution in the coherent regime, we numerically simulated the theoretical models in Eqs.~\eqref{eq:ran_fields} and \eqref{eq:Pnzeta}.  The first task is to determine the parameters $\bm{\zeta}$ in Eq.~\eqref{eq:ran_fields} and the (inverse) operating temperature $\beta$ in Eq.~\eqref{eq:Pnzeta}.  We use the fast anneal data in the coherent regime to obtain an estimate for $\bm{\zeta}$, and both fast and standard anneal data in the fully thermalized regime to estimate $\beta$.  Since the theoretical models given by Eqs.~\eqref{eq:ran_fields} and \eqref{eq:Pnzeta} do not have closed form analytical solutions, we calculate the model numerically using $\bm{\zeta}$ and $\beta$ as free parameters to fit to the experimental data.  This requires a large number of Monte Carlo samples to obtain sufficiently accurate data, which would be impractical for larger systems.
The simulation data are also post-selected for single domain walls in the same way as the experimental data, resulting in a similar reduction in total samples.

\subsubsection{Random field strength fitting}\label{ssec:zeta_fit}

In order to fit the strength of the random fields, closed-system quantum simulations were fitted to the experimental data with a $0.005\,\mu$s anneal time, as these are expected to show almost entirely coherent behaviour. In the fast anneal regime, the pinning of spins at each end is achieved with flux biases. The Hamiltonian used for the simulations is therefore modified from Eq.~\eqref{eq:H_res_fields} to
\begin{equation}\label{eq:H_flux_biases}
\begin{aligned}
H(t)=&-\frac{A(s)}{2}\sum_{i=1}^N\sigma_i^{x}  + \frac{B(s)}{2}H_{\text{chain}} \\
&+\frac{\sqrt{B(s_c)B(s)}}{2}H_{\text{bias}} + H_{\text{fields}},
\end{aligned}
\end{equation}
where $s_c$ is the point where $B(s_c) = A(s_c)$ in \cite{DWave_schedule_docs}. 

Fitting the simulation to the experimental data at $0.005\,\mu$s in the fast anneal protocol, with the disorder strength $\bm{\zeta}$ as a free parameter (for details see Sec~(\ref{ssec:numsim}), we obtain $\bm{\zeta} = 0.101 \pm0.007$ GHz. 
Comparison of the fit with the experimental data is shown in Fig.~\ref{fig:fast_anneal_sigma}.

\subsubsection{Operating temperature fitting}\label{ssec:beta_fit}

%
\begin{table}[tb!]
\centering
\renewcommand{\arraystretch}{1.15}%
\begin{tabular}{|l|c|c|c|c|c|}
\hline
\multicolumn{2}{|c|}{} & L & Time($\mu$s) & $c_{fit}\pm\Delta c_{fit}$ & Calculated $T\,[\mathrm{mK}]$ \\ \hline
\multirow{2}{*}{\textbf{}}
    & \multirow{1}{*}{Standard} & 10   & 20   & 0.278 $\pm$ 0.055 &  $17.46 \pm 3.67$  \\ \cline{2-6}
    & \multirow{1}{*}{Fast}    & 12 & 20 & 0.276 $\pm$ 0.042 & $17.59 \pm 2.92$  \\ \cline{2-6} 
    & \multicolumn{3}{c|}{Average} & 0.277 $\pm$ 0.034 & $17.53 \pm 2.34$ \\ \cline{1-6}

\hline

\end{tabular}
\caption{Values of dimensionless $c_{fit} = \beta\bm{\zeta}$ are extracted by fitting the experimental data to the numerically calculated Boltzmann distribution at 20 $\mu$s in the standard anneal protocol and fast anneal protocol respectively. 
 Here, $\bm{\zeta}=0.101\pm0.007 \ GHz$ is the magnitude of the disorder fields expressed in GHz, and $\beta=h/(k_BT)$ is the inverse temperature in frequency units, expressed in \(\mathrm{GHz}^{-1}\). The temperature of the device is estimated using $T=\frac{h\zeta}{k_Bc_{\mathrm{fit}}}$. Uncertainty, $\Delta c_\text{fit}$, is the standard deviation of the $c_\text{fit}$ values obtained by fitting each of $100,000$ bootstrap resamples of the experimental data to the numerically calculated Boltzmann distribution, and is propagated to the estimated temperatures.}
\label{tab:fits}
\end{table}

\begin{figure}[tb!]
 \includegraphics[width=0.95\linewidth]{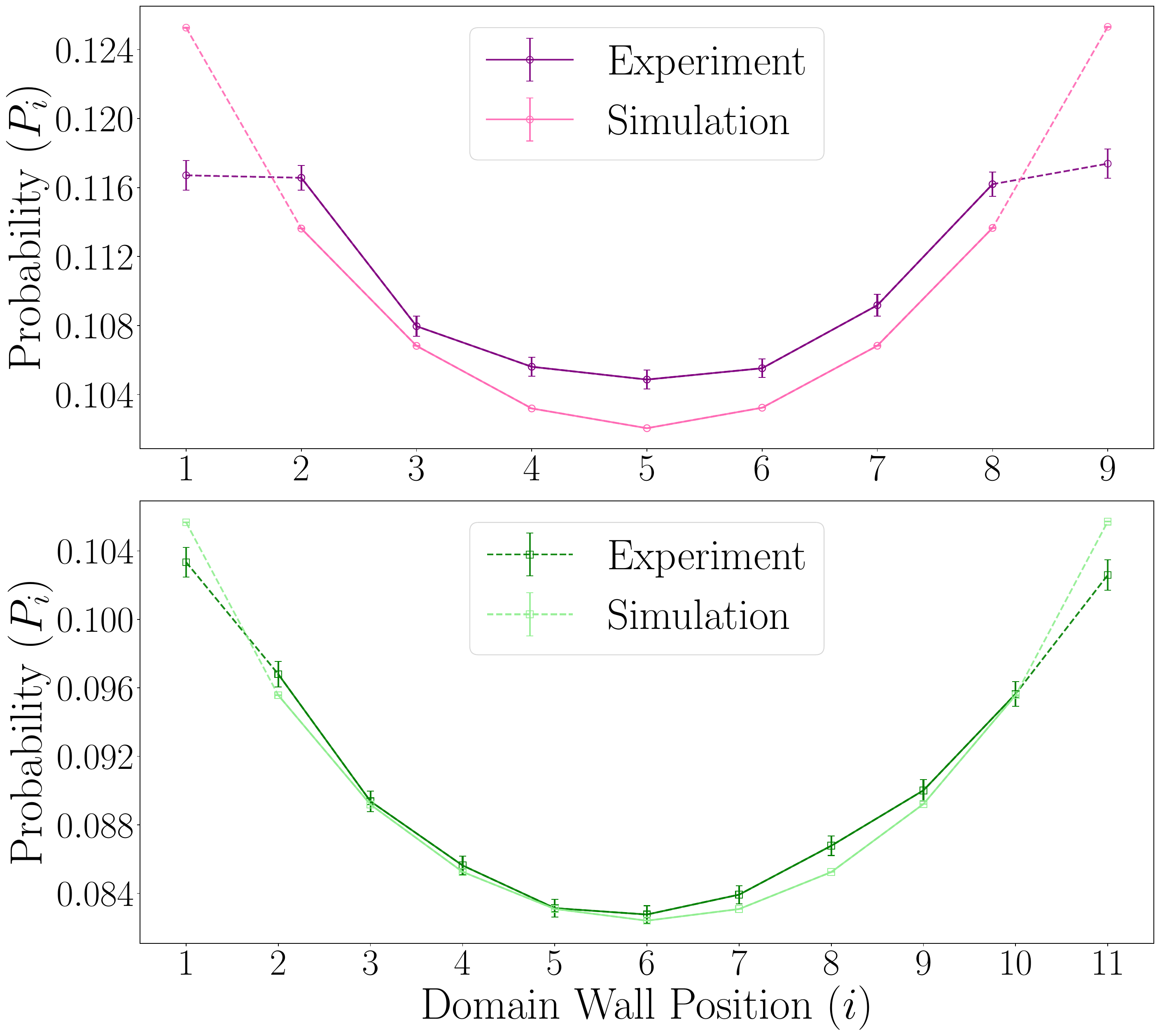}
\caption{\label{fig:normal_fast_single_plot} Standard (top) and fast (bottom) anneals for an anneal time of $20\,\mu$s, experimental data (solid lines) and simulations (dotted lines) fitted to extract the best fit values of $\beta\bm\zeta$, of $0.278 \pm 0.055$ and $0.276 \pm 0.042$ for the standard and fast anneal protocols, respectively. Terminal sites are excluded from the fit, consistent with the procedure in Ref.~\cite{Chancellor_2022}. Error bars in simulated distributions represent standard errors. The faded curve show the data with terminal sites included. In both protocols, all experimental data are averaged across multiple chains with $10^4$ experimental runs. Error bars in the experimental data represent the uncertainty obtained by combining in quadrature the propagated binomial uncertainty from finite domain wall occurrences and the standard error of the mean across chains.}
\end{figure}

For standard anneals of $20\,\mu s$ duration, the experimental data is expected to show fully thermal behaviour matching a Boltzmann distribution as in Eq.~\eqref{eq:Pnzeta}.  In Eq.~\eqref{eq:Pnzeta}, the free parameter for fitting is $\beta \bm{\zeta}$, so we cannot independently estimate both from the standard anneal data and must use the value of $\bm{\zeta}$ estimated from the fast anneal data to estimate $\beta$.  Using $c = \frac{1}{k_BT} \times \bm{\zeta}$ together with relations $\nu = k_B T/h$ where $\nu$, $T$, $k_B$ and $h$ represent frequency, temperature of annealer, Boltzmann constant and  Planck’s constant respectively, we solve for $\bm{\zeta}$ in units of $GHz$. 
The results of this fitting are shown in table \ref{tab:fits},
and the plots corresponding to the fitted values of $\beta \bm{\zeta}$ are shown alongside the experimental data in Fig.~\ref{fig:normal_fast_single_plot}.
It can be seen in table \ref{tab:fits} that the values from the standard and fast anneals are in good agreement with each other, despite the fit for the standard anneal being significantly less good in Fig.~\ref{fig:normal_fast_single_plot} (top).

Our temperature estimate, averaged across the estimate values obtained from the standard and fast anneal protocols, is $17.53 \pm 2.34$ which is within is within one standard deviation of the reported value of $15.4 \,$mK, of the D-Wave Advantage 4.1 annealing device.

\subsection{Understanding domain wall behavior}
\label{ssec:sim_fast_anneal}

The two factors in our models that influence the non-ideal spin chain dynamics are the random fields of strength $\bm{\zeta}$, present at all timescales, and the thermalization that degrades the initial coherent dynamics at longer anneal times, due to the finite operating temperature $\beta$.
Having used the fastest and slowest anneal times to obtain values for the two parameters $\bm{\zeta}$ and $\beta$, we now analyse the data for the anneal time of $0.01\,\mu$s to determine the relative influence of both factors.  As a baseline, we also simulated the ideal closed system without random fields or thermal effects.

\begin{figure*}[tb!]
\includegraphics[width=0.99\linewidth]{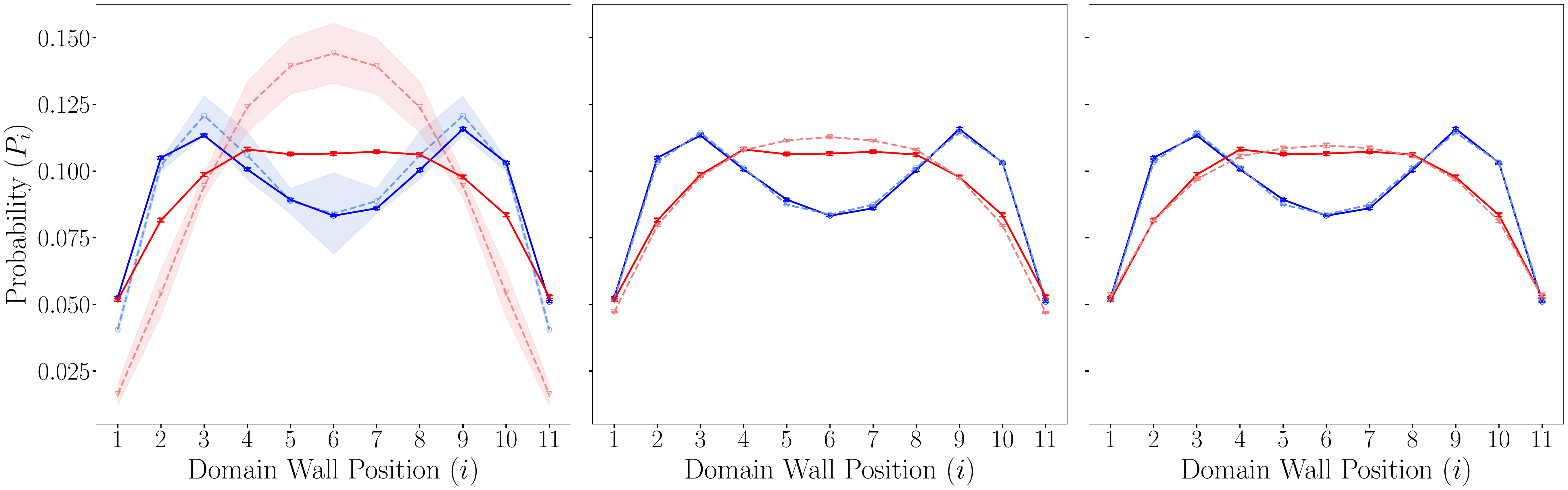}
\caption{\label{fig:fast_anneal_sigma}
Probability distributions from fast anneal experiments (solid lines) and the corresponding numerical simulations of coherent evolution (dashed lines) at $0.005,\mu$s (blue) and $0.01,\mu$s (red). The panels show simulated coherent evolution according to Eq~\eqref{eq:H_res_fields} with $\bm{\zeta}=0$ (left), simulated coherent evolution with residual random fields, of bias strength, $\bm{\zeta} = 0.101 \pm0.007 \ GHz$, based on Eq~\eqref{eq:H_res_fields} (middle) and the same latter simulation mixed with a thermal Boltzmann distribution as in Eq~\eqref{eq:mix} (right) with mixing parameters $b\approx 0 $ with a $95\%$ percentile confidence interval of [0.000,0.056] at $0.005\,\mu$s, and $b = 0.105 \pm 0.025 $ at $0.01\,\mu$s. 
The shaded regions (left) show the $\pm 1\sigma_{SD}$ standard deviation of the simulated probabilities around $t$, averaged over 10,000 runs, with  $\sigma_{t}=0.05t\,\mu$s. Error bars in the simulations represent standard errors. Error bars in the experimental data represent the uncertainty obtained by combining in quadrature the propagated binomial uncertainty from finite domain wall occurrences and the standard error of the mean across chains.}
\end{figure*}

\subsubsection{Ideal dynamics}
\label{ssec:sim_fast_anneal1}
For our first set of simulations, we use Eq.~\eqref{eq:HTI}, the ideal closed quantum system simulation without random fields or thermal effects, and with $H_{\text{problem}} = H_{\text{chain}}$ from Eq.~\eqref{eq:Hspinchain}. Our simulations have strong oscillatory profiles, and in order to smooth this out, we average the simulated probability distribution at each anneal time over Monte Carlo realizations with the anneal time $t$ sampled as $t+\delta t$, where $\delta t \in \mathcal{N}(0,\sigma_t^2)$ with $\sigma_t=0.05t$. The simulation results for the fastest anneal times, $0.005\,\mu$s and $0.01\,\mu$s , are shown in Fig.~\ref{fig:fast_anneal_sigma} (left) along with the probability distributions observed in the experiment. The shaded areas show one standard deviation in the numerical distributions.  For the fastest anneal time of $0.005\,\mu$s, the experiments are close to one standard deviation from the ideal distribution, showing that the effects of the random fields are relatively small at the shortest time.  In contrast, for $0.01\,\mu$s anneal time the simulations are very far from the experimental data.

\subsubsection{Random fields only}
\label{ssec:sim_fast_anneal2}
We then added the random fields, Eq.~\eqref{eq:H_flux_biases}, using $\bm{\zeta} = 0.101 \pm0.007$ GHz obtained from the fitting procedure, which means the simulations match the shortest anneal time used for the fitting as closely as possible. For an anneal time of $0.01\,\mu$s, the correspondence with the experimental data is pretty good, but not perfect, as can be seen in Fig.~\ref{fig:fast_anneal_sigma} (middle), suggesting that some effects from thermalization are already present.

\subsubsection{Random fields plus thermalization}
\label{ssec:sim_fast_anneal3}

To introduce a small amount of thermalization into the simulations, we used a linear combination of the random fields and thermal distributions,
\begin{equation}
    P_{\text{mix}} = (1 - b)P_{\text{coherent}} + b P_{\text{thermal}}
    \label{eq:mix}
\end{equation}
where $P_{\text{thermal}}$ is the averaged Boltzmann distribution, 
and $P_{\text{coherent}}$ is the distribution obtained from the simulations with the random fields Eq.~\eqref{eq:H_res_fields}. 

We take the thermal probability distribution obtained from our experimental data at $20\,\mu$s using the fast anneal protocol, see Fig.~\ref{fig:fast_anneal} (bottom).
The mixing weight $b$ between the coherent quantum distribution and the thermal distribution is optimized by minimizing the Kullback–Leibler (KL) divergence between the experimental data and the numerical distribution $P_{\text{mix}}$ from Eq.~\eqref{eq:mix}
\begin{equation}
    D_{\mathrm{KL}}\!\left( P_{\mathrm{exp}} \,\middle\|\, P_{\mathrm{mix}} \right) 
     = \sum_{j = 0}^{N} P_{\mathrm{exp}}(j) \, \ln \frac{P_{\mathrm{exp}}(j)}{P_{\mathrm{mix}}(j)}.
\end{equation}

Fig.~\ref{fig:fast_anneal_sigma} (right) shows that the numerically simulated data now agree pretty well with the experimental results for $0.01\,\mu$s, with $b = 0.105 \pm 0.025 $ indicating an approximately $10\%$ contribution from thermal fields in the presence of random fields. 
Adding a small amount of thermal distribution to the distribution for an anneal time of
$0.005\,\mu$s does not significantly improve the already very good fit, we find $b\approx 0 $ with a $95\%$ percentile confidence interval given by its  $2.5\%$ and $97.5\%$ percentiles of [0.000,0.056] suggesting thermalization effects have not started to take effect yet.

The uncertainty in the fitted values of $b$ is estimated through generating multinomial bootstrap samples from the experimental distribution and refitting b for each sample, giving the reported error as the standard deviation of the resulting bootstrap distribution of $b$. 

\begin{figure*}
\includegraphics[page=1,width=0.9\linewidth]{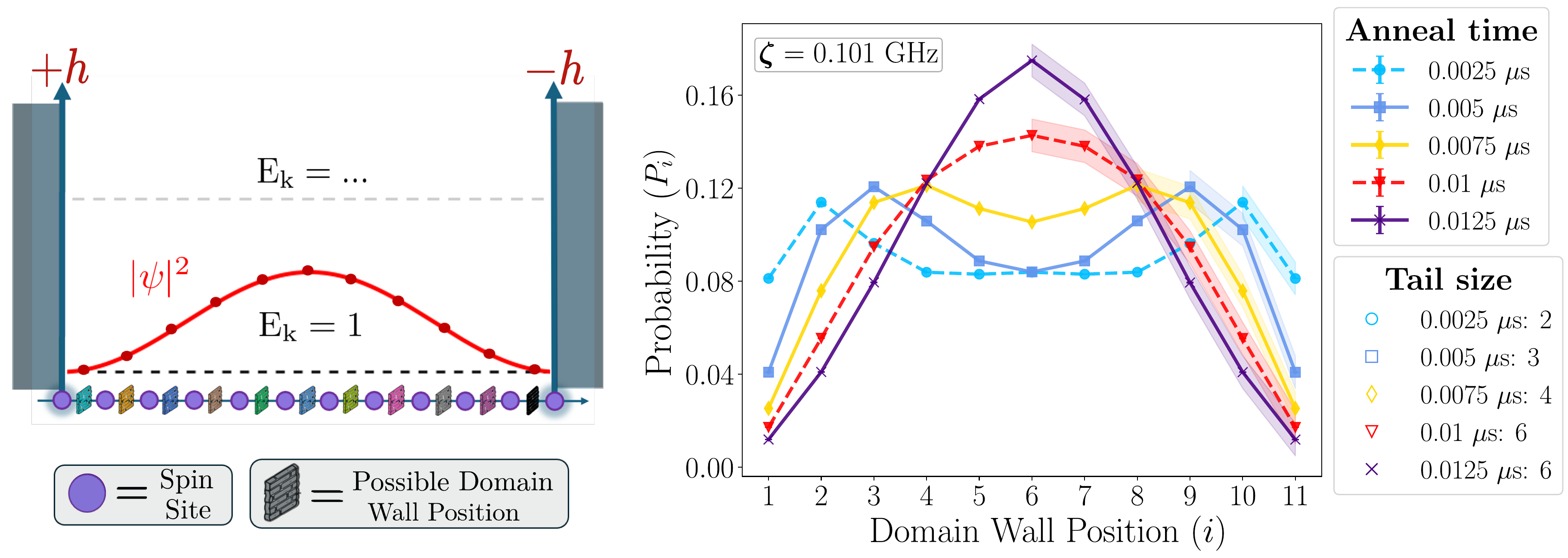}
\caption{\label{fig:tail} Left: schematic of the particle-in-a-box showing the $E_k = 1$ ground state sine squared probability density, with the corresponding spin chain shown along the $x$ axis, where the boundary spins are pinned by the $h$ fields. Right: Simulation using model~\textrm{II} , with $100\,000$ simulation runs, of the domain wall probability distributions for different anneal times, with a static field strength of $\bm{\zeta}=0.101 \pm 0.007$ GHz (right). The shaded part illustrates the idea of the tail, which is the number of data points that monotonically decrease on either side from the middle. The tail size incorporates the number of these data points on one side. Simulations were averaged over $100\,000$ runs with error bars representing standard error.}
\end{figure*}

\section{A benchmarking metric from the domain wall distribution tail}
\label{sec:tail metric}

As we have seen, at sufficiently fast anneal times, the domain wall distribution on a D-Wave System approximately reproduces the expected sine-squared profile of a particle in its ground state for a fully coherent quantum annealer. In contrast, for anneal times that are too short or too long, the distribution dips in the centre.  We therefore propose to ue the tails, defined as the number of domain walls for which the distribution increases monotonically from the pinned ends towards the centre, to quantify how closely the measured profile approaches this ideal form. For an ideal sine-squared distribution, this is half of the chain length.  For the U-shaped thermal distribution with random fields, this is zero.  We can therefore normalize by half of the chain length to obtain a metric between zero (thermalized) and one (fully coherent). For very fast anneals, when the single domain wall particle-in-a-box distribution has not yet been reached, the tails are not zero, but are definitely less than one, for example, $3/6 = 0.5$, as seen in our experimental data for $0.05\,\mu$s in Fig.~\ref{fig:fast_anneal}.  For our experimental data at $0.1\,\mu$s the tail length is five, which is $5/6$ of the maximum possible in an $L=12$ chain, indicating the system is not perfectly coherent.  For $0.3\,\mu$s and $20\,\mu$s the U-shaped distributions have tail length zero. In Fig.~\ref{fig:tail} (right), we show using simulations how the metric varies for a fully coherent system with random fields at the strength we measured in our D-Wave experiments.  It clearly shows the chain length growing as the domain wall distribution reaches equilibrium.
 
\begin{figure*}[t]
\includegraphics[width=0.95\linewidth]{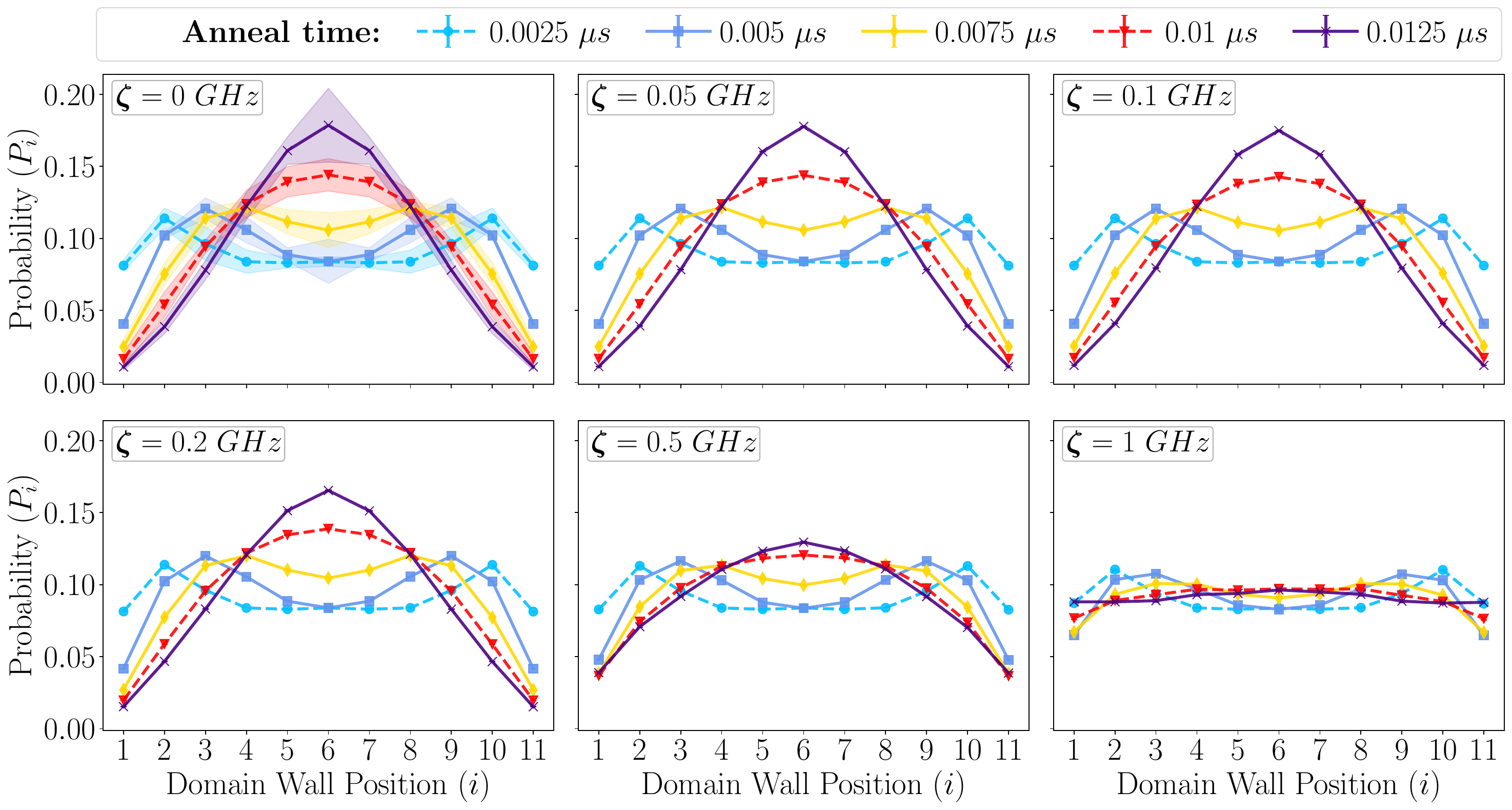}
\caption{\label{fig:noise_anneal} Simulated domain wall distributions using random fields but no thermal effects, for different anneal times and noise strengths. Simulations were run for $100\,000$ Monte Carlo samples. Error bars represent standard error. 
 } 
\end{figure*}

\begin{figure}
\includegraphics[width=0.92\linewidth]{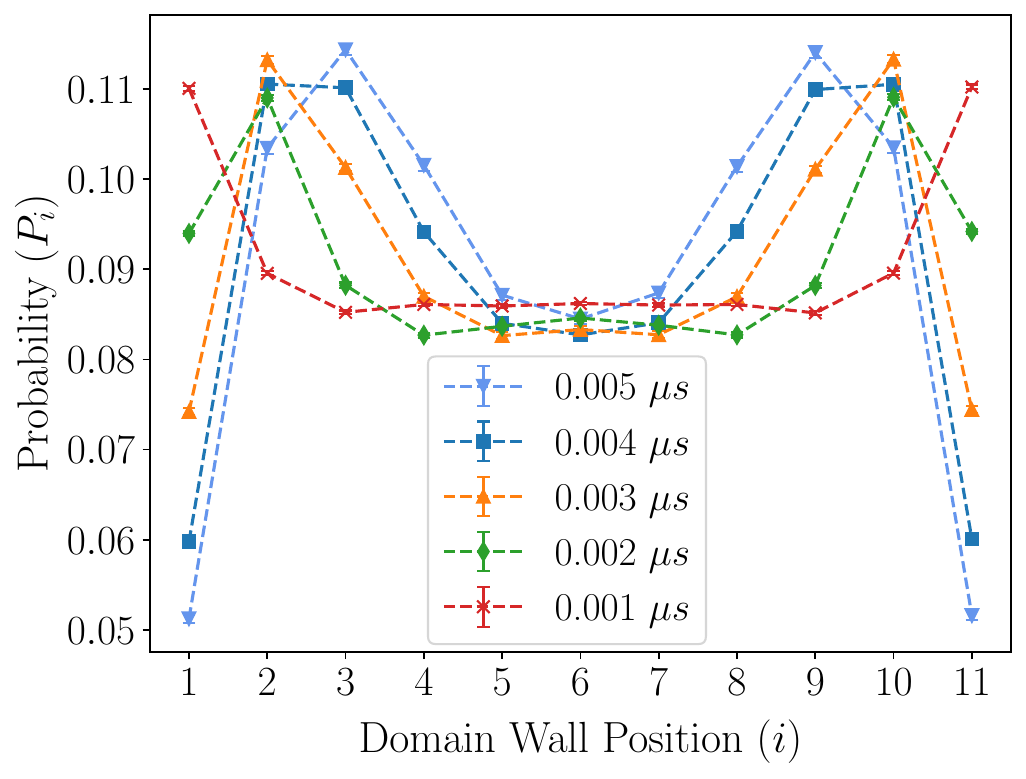}
\caption{\label{fig:outside} Domain wall probability distributions using at ultra short anneal times. Simulations for anneal times $ T < 0.005\,\mu$s, down to $ 0.001 \,\mu$s, with the  $0.005 \,\mu$s for comparison.  Simulations were run for $10^4$ repeats using $\bm{\zeta}= 0.101 \pm 0.007$ GHz. Error bars represent standard error. 
}
\end{figure}

Figure \ref{fig:noise_anneal} shows simulations of the distribution for different noise strengths and anneal times, still all in the coherent regime.  For larger values of $\bm\zeta$, the distributions flatten, but retain the rising tail metric signatures until $\bm\zeta$ reaches values comparable with the $h$ and $J$ values in the problem Hamiltonian, where the device is too noisy to be useful to instantiate a problem accurately.
Fig.~\ref{fig:outside} shows simulations for anneal times below $0.005\,\mu$s, a regime that is currently inaccessible on D-Wave hardware. These simulations provide insight into the expected annealing behavior in this ultra fast annealing limit.  For the fastest simulation at $0.001\,\mu$s, the domain wall has not yet localized significantly, while for $0.002\,\mu$s through $0.004\,\mu$s the partial localization gives a metric value of 2/6.

\section{Discussion and Conclusion}\label{sec:discuss}

We have shown that the fast anneals now available on D-Wave Systems are sufficiently coherent to produce a good approximation of the ideal sine-squared particle-in-a-box distribution predicted in \cite{Chancellor_2022}.  Our simulations suggest there is about $10\%$ thermal effects in the distribution measured on a D-Wave Advantage 4.1 system for anneal times of $0.01\,\mu$s.

Our experiments (at $0.005\,\mu$s and $0.01\,\mu$s), supported by simulations, suggest that the extent of the domain wall tail provides a simple and robust metric for benchmarking analogue quantum devices. The tail morphology is a result of quantum interference and therefore reveals the degree of delocalization attainable for different anneal times. In ideal coherent evolution, the interference pattern can extend to the center of the chain, producing a characteristic sine-squared profile. At sufficiently long anneals, noise and decoherence shorten this interference range, while very fast anneals prevent the wavefunction from fully spreading, leading to a truncated tail. 

We therefore propose the tail extent as a device agnostic benchmark, suitable not only for superconducting flux qubit annealers but also for emerging analogue quantum simulators such as neutral atom Rydberg platforms.  This extends the proposed benchmarks for quantum annealing platforms in \cite{lall2025reviewcollectionmetricsbenchmarks}.
However, Rydberg errors are dominated by loss -- future work will determine whether this metric provides information not easily obtainable through other metrics. As a next step, we will use this metric to benchmark Rydberg platforms. This will allow us to further test the effectiveness of this metric while also providing a broader performance comparison between two different quantum computing platforms.

\begin{acknowledgments}
SC is funded by the International Graduate School in Quantum Technologies (IGSQT). SC thanks Daniel Walker for helpful advice on data management. The authors thank Axel Daian for useful discussions on the annealing protocol which the annealing devices use.  AH is funded by UKRI EPSRC PhD studentship number 2745408. NC and VK funded by UKRI EPSRC grants EP/T026715/2, EP/T001062/1, EP/W00772X/2, and EP/Y004566/1, EP/Z53318X/1, and the UKRI DRI and STFC funded CCP-QC Bridge Project -- extended case studies for neutral atom hardware.
\end{acknowledgments}

%


\appendix

\section{Chain embedding}
\label{app:emb}

\begin{figure}[!tb]
\includegraphics[width=1\linewidth]{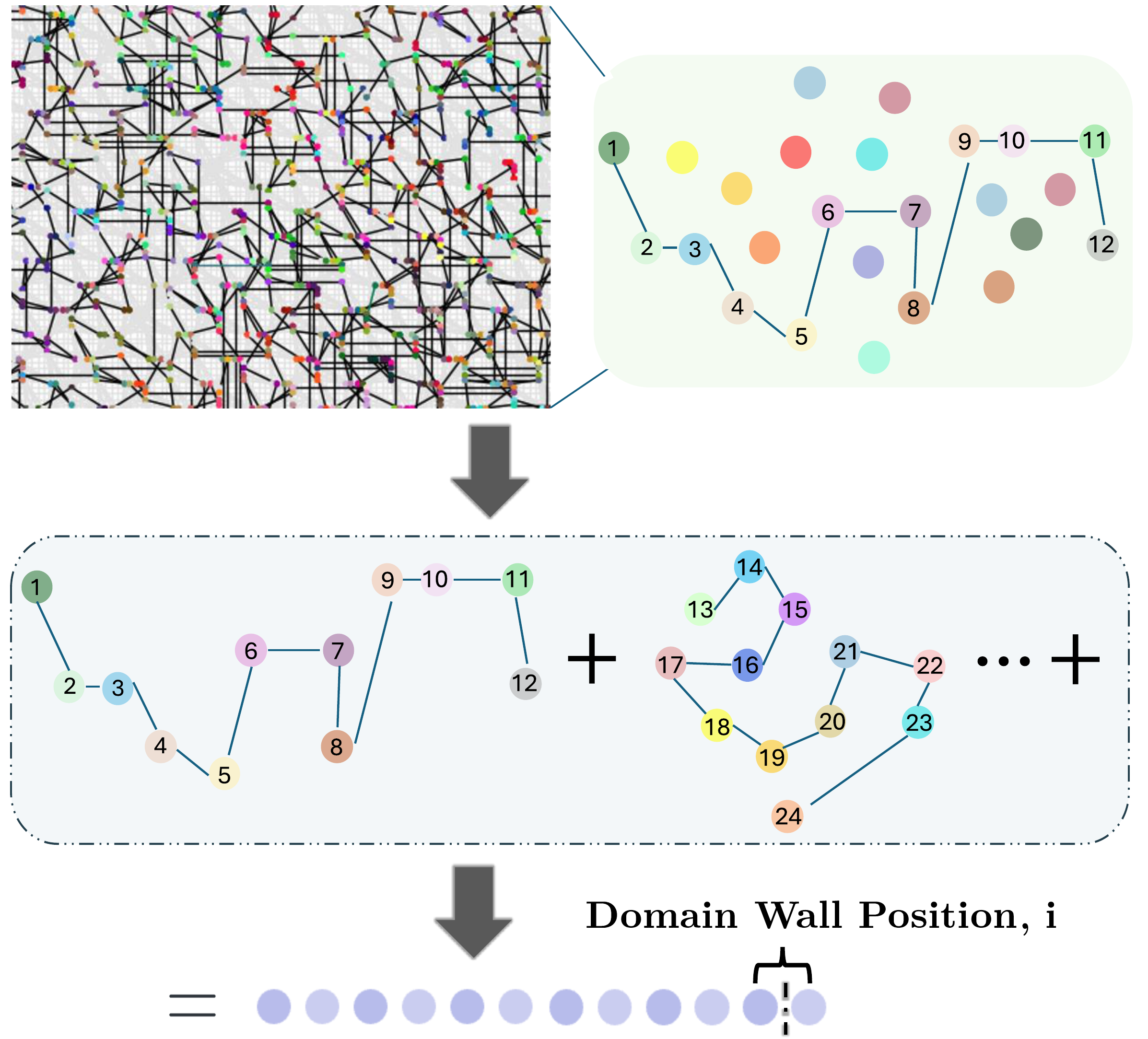}
\caption{\label{fig:map1} Left shows a segment of the Pegasus hardware graph and on the right shows a simplified chain from from the graph using one to one embedding.}
\end{figure}
In order to make full use of the D-Wave hardware qubits, we fitted many chains onto the chip, placed randomly to avoid systematic errors arising from the graph structure.  Fig. \ref{fig:map1} illustrates this on a segment of the Pegasus hardware graph.

We used a path construction algorithm based on Warnsdorf’s rule \cite{ganzfried2004simple}, originally introduced as a heuristic for solving the knight’s tour problem, to traverse the connectivity graph from a random starting point. Starting from a chosen vertex, the algorithm repeatedly extends the path by selecting the connected vertex with the lowest degree, and then removing the chosen vertex from the working graph for the remainder of the algorithm. Prioritizing vertices of low degree can be seen as removing the least important vertices of the graph, preventing the structure from becoming disconnected to the extent that is possible.

The algorithm proceeds in two phases. It first extends the path forward until reaching a vertex with no remaining neighbors and then reverses the constructed path and continues from the opposite end. This forward–reverse procedure ensures that the final path terminates with dead ends at both ends, thereby maximizing coverage. The process terminates once no further unvisited vertices remain. We repeat runs of this algorithm from random starting qubits until the resulting path covers at least $99.8\%$ of the hardware graph, ensuring near complete qubit utilization. This typically takes $<10$ attempts for the Pegasus graph. 

\section{Gauge averaging}
\label{app:gauge}

\begin{figure}
\includegraphics[width=1\linewidth]{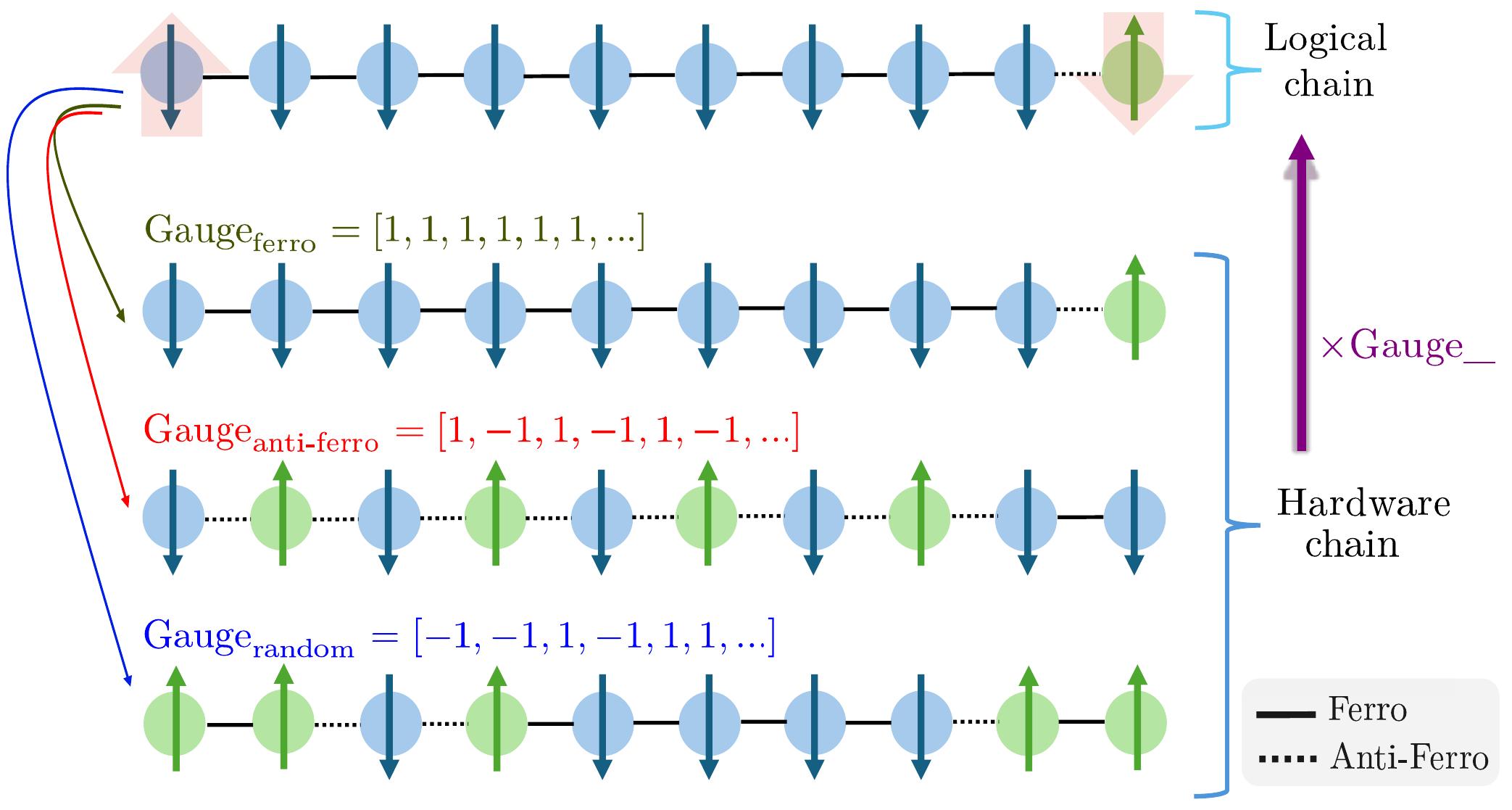}
\caption{\label{fig:gauge} The logical problem (top row) a ferromagnetic chain with all couplers set to $J=-1$, with end spins pinned by opposite local fields $h=\pm 2$ (left/right). Various gauge transformations are shown in the lower rows.}
\end{figure}

%
Fig.~\ref{fig:gauge} illustrates the idea of \emph{gauge averaging}, where the same logical problem is run under different gauge choices for each qubit. Gauge transformations are a standard technique for averaging out systematic errors due to variations in the qubits across the chip. Gauge averaging is necessary because real QPUs suffer from calibration errors and fabrication imperfections, which cause spins and couplers to respond differently under different spin orientations. The averaging minimizes these asymmetric errors.  

\begin{figure}[tbh!]
\includegraphics[width=1\linewidth]{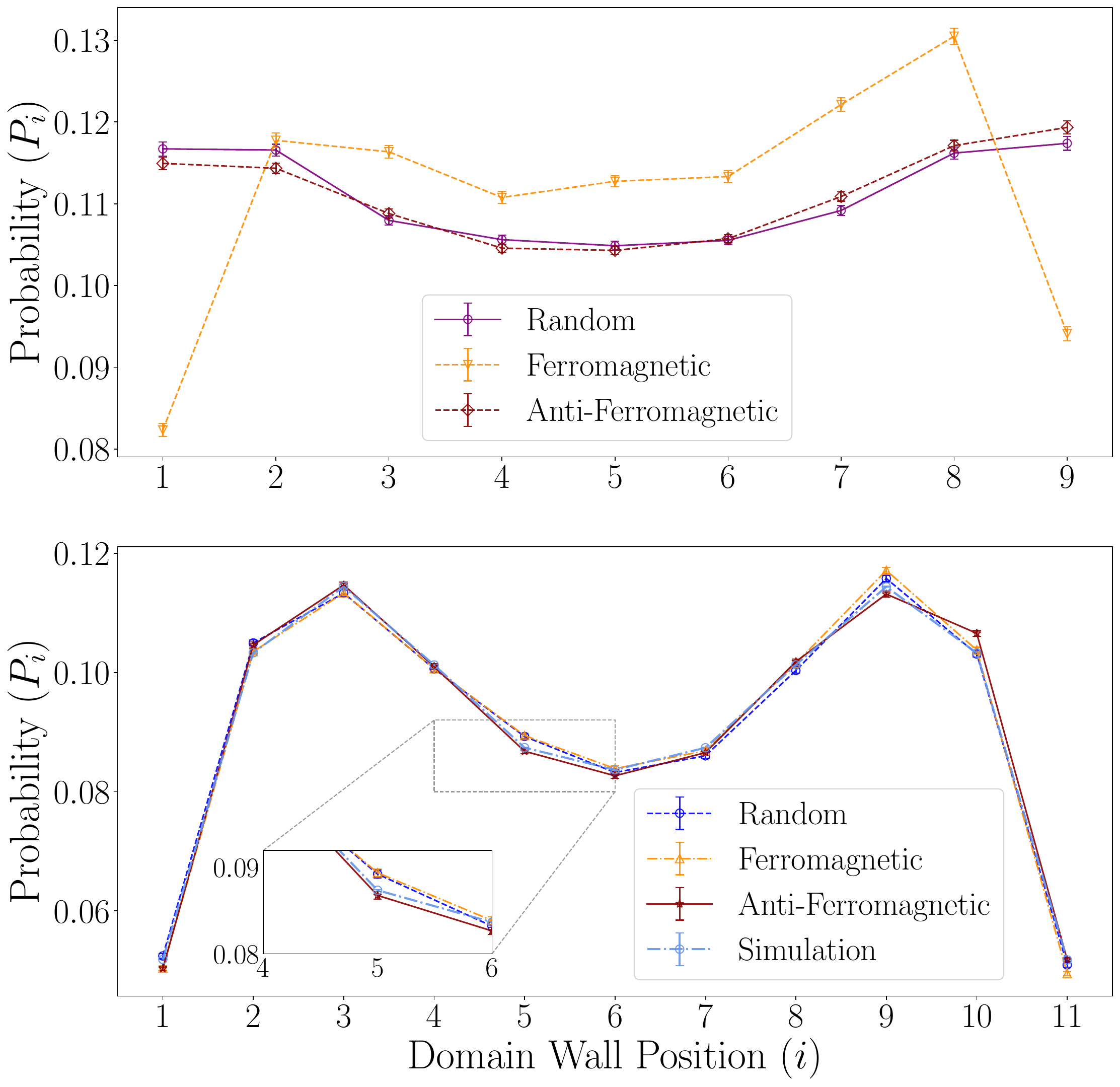}
\caption{\label{fig:Gstrength} 
Probability distribution for standard anneal protocol (top) for an anneal time of $20\,\mu$s for chain length 10 with data averaged over random gauges (circles),  ferromagnetic gauges (inverted triangles), equivalent to applying no gauge, and antiferromagnetic gauges (diamonds). Fast anneal protocol (bottom) for $0.005\,\mu$s for chain length 12 with data averaged over random gauges (circles), ferromagnetic gauges (triangles), and antiferromagnetic gauges (stars). In both protocols, all data are averaged across multiple chains with $10{,}000$ experimental runs. Error bars in the experimental data represent the uncertainty obtained by combining in quadrature the propagated binomial uncertainty from finite domain wall occurrences and the standard error of the mean across chains. Simulations are shown at $0.005\,\mu$s with $\bm{\zeta = 0.101 \pm 0.007}$ (for details see ~\ref{ssec:numsim}). Simulation at $0.005\,\mu$s confirms that the data do not change with gauge in the fast anneal protocol (see inset) with error bars representing standard error.}
\end{figure}

Our results for the standard anneal protocol are shown in Fig.~\ref{fig:Gstrength} (top). We observe gauge-dependent differences in the depth and symmetry of the U-shaped domain wall distributions, indicating that the couplers influence the effective local fields $\zeta_i$. The ferromagnetic gauge produces the shallowest profile and exhibits more pronounced edge asymmetry than the random and antiferromagnetic gauges. 
Our findings are opposite to that reported in \cite{Chancellor_2022} on an earlier generation of D-Wave hardware, where the ferromagnetic gauge produced the deepest U-shaped distribution and was attributed to stronger coupling to noise. Such differences are plausible across hardware generations, where calibration procedures and hardware level biases may differ. 
In both studies, however, the gauge dependence indicates that because of systematic hardware effects, a single gauge does not fully reproduce the gauge averaged distribution.  The ferromagnetic gauge is particularly affected in our data, exhibiting a noticeable asymmetry that may arise because the boundary pinning has the same orientation across all chains, with no gauge randomization to average out systematic biases.

In the fast anneal runs at $0.005\,\mu$s, all three gauge choices produce nearly identical domain wall probability distributions. In contrast to the standard anneal, where the distribution depends noticeably on the gauge, this indicates that gauge dependent hardware biases have much less influence at the shortest anneal time. Such gauge independent behaviour is consistent with predominantly coherent dynamics, since at very short anneal times gauge dependent hardware biases have less time to influence the evolution. 
The M-shape arises as a result of the anneal being sufficiently fast that the domain wall does not have enough time to fully delocalize across the chain, so the wave function does not have enough time to develop the standing wave profile across the chain as discussed in Sec.~\ref{ssec:fast_anneal}. The closed system simulation at $0.005\,\mu$s reproduces this behaviour, supporting the interpretation that the profile originates from early time coherent evolution of the domain wall wave function. In the ideal closed system simulation, gauge transformations are equivalent because gauge dependent biases from the hardware do not play a role.

\section{Calibration and control errors} \label{app:single_error}

\begin{figure}
\includegraphics[width=1\linewidth]{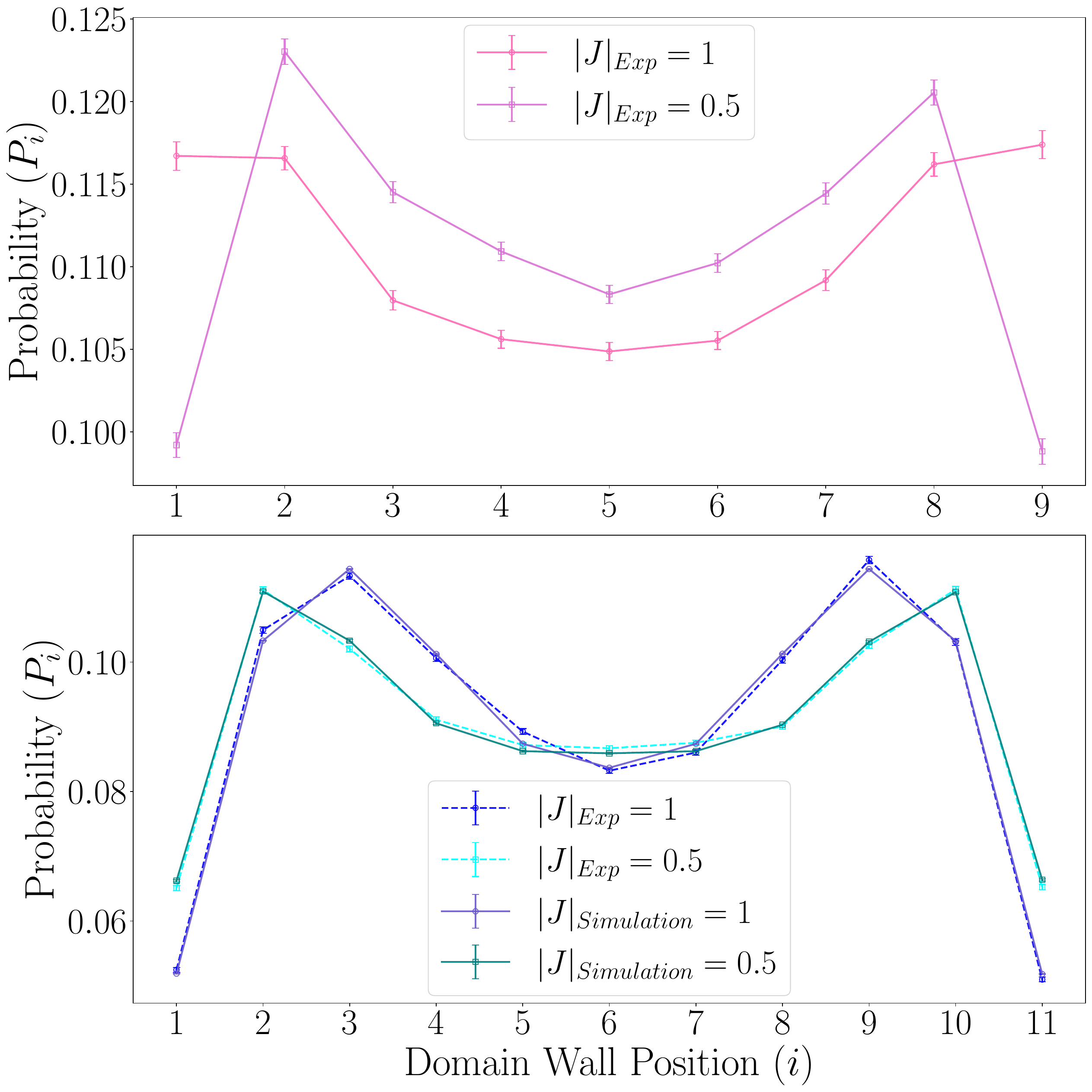}
\caption{\label{fig:Jstrength} 
Standard anneal protocol (top) for an anneal time of $20\,\mu$s for chain length 10 with  couplings, $J =1$ (circles) and $J =0.5$ (squares).  Fast anneal protocol (bottom) showing the same (dashed lines) for an anneal time of $0.005\,\mu$s for chain length 12 with the addition of simulation for $J =0.5$ (solid line). In both protocols, all data are averaged across multiple chains with $10^4$ experimental runs. Error bars in the experimental data represent the uncertainty obtained by combining in quadrature the propagated binomial uncertainty from finite domain wall occurrences and the standard error of the mean across chains. Simulation at $0.005\,\mu$s using $J=1$ were carried out, with error bars representing standard error. Simulations are shown at $0.005\,\mu$s for $J =1$ with $\bm{\zeta = 0.101 \pm 0.007}$ (for details see \ref{ssec:numsim}) and using the same fitted $\bm{\zeta}$, simulations were run for for $J =0.5$ using $10^5$ repeats. Error bars in simulation show the standard error.)}
\end{figure}

A ferromagnetic bond has energy $-J$ when neighbouring spins are aligned and $+J$ when they are anti-aligned. Therefore, the energy required to transition from an aligned state to one containing an additional domain wall sets the excitation gap, with a cost of $(+J-(-J)=2J)$. Halving (J) halves the energy cost of creating a domain wall, narrowing the excitation gap and making domain wall excitations energetically easier. In the fast anneal protocol for a chain of $12$ spins, this places the system in a Landau Zener regime \cite{King:2022phl}, where excitations arise from diabatic transitions across the minimum gap and the transition probability depends exponentially on $(\Delta^2/v)$ \cite{PhysRevLett.95.245701}, where the minimum energy gap $\Delta$ scales with the coupling strength and $v$ is the rate at which the system is driven through the avoided crossing. We can indeed see from table~\ref{tab:domain_count}, the lowest number of single domain wall configurations, $5145.4 \pm 7.6$, and the highest number of excited domain wall configurations, $20197.6 \pm 16.0$, occur when $J$ is halved compared to single domain wall configurations, $7072.9 \pm 11.2$, and the excited domain wall configurations, $15893.6 \pm 17.8$, when $J=1$. 
To pin the boundary spins, the applied field is set by the relative strength $h/J$; therefore, halving $J$ while keeping $h$ fixed increases $h/J$, making the boundary spins more strongly pinned relative to the bulk. However, this may enhance edge effects in the domain wall distribution probability distribution as is suggested by Fig.~\ref{fig:Jstrength}. As reported in \cite{Chancellor_2022}, the average domain wall probability distribution probes effective random longitudinal fields, while the averaging suppresses coupler dependent error contributions. We similarly find little change in the standard anneal profile when $J$ is reduced, whereas the fast anneal profile changes more noticeably because $J$ directly affects the short time coherent dynamics as we discussed. As the Hamiltonian scales the programmed Ising terms through $B(s)$, the hardware is subject to flux noise and intrinsic control errors in both local fields and couplers \cite{DWaveAnnealing_control}. Thus, in weaker $J$, disorder fields can have a stronger relative influence on domain wall dynamics since $\bm{\zeta}/J$ becomes larger and may suppress coherent spreading and produce stronger spatial localization, similar to Anderson localization \cite{Karamlou2021QuantumTA}.

\section{Post-selecting single domain walls}
\label{app:post_select}
\begin{figure*}
\includegraphics[width=0.85\linewidth]{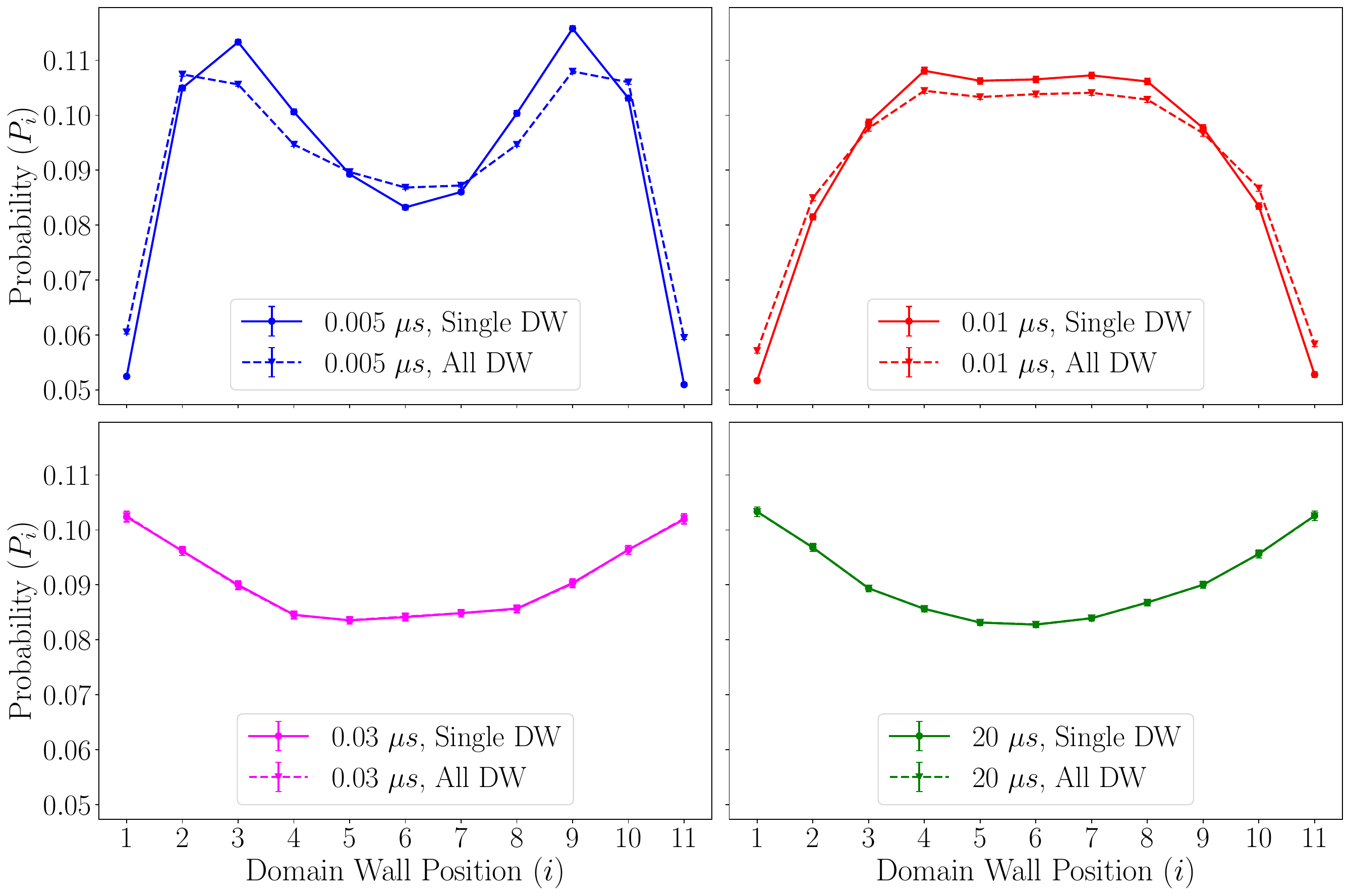}
\caption{\label{fig:domain_multi_wall} Probability distribution from the fast anneal protocol for a chain length of 12 with $10\,000$ runs for different anneal times showing all domain wall instances and single domain wall instance. For showing all domain walls, the raw data is used and for  the single domain wall instances, we post select the raw data. On average across multiple chains per run this leaves all domain wall count at $15893.6 \pm 17.8$ and single domain wall count at $7072.9 \pm 11.2$ for $0.005\,\mu$s, all domain wall count at $11753.1 \pm 19.1$ and single domain count wall at $9119.4 \pm 17.4$ for $0.01\,\mu$s, all domain wall count at $10061.1 \pm 26.8$ and single domain wall count at $9960.9 \pm 26.7$ for $0.03\,\mu$s, and all domain wall count at $9792.6 \pm 21.3$ and single domain wall count at $9762.9 \pm 21.3$ for $20\,\mu$s (see table~\ref{tab:domain_count}).Error bars in the experimental data represent the uncertainty obtained by combining in quadrature the propagated binomial uncertainty from finite domain wall occurrences and the standard error of the mean across chains.}
\end{figure*}

\begin{figure*}[!tb]
\includegraphics[width=1\linewidth]{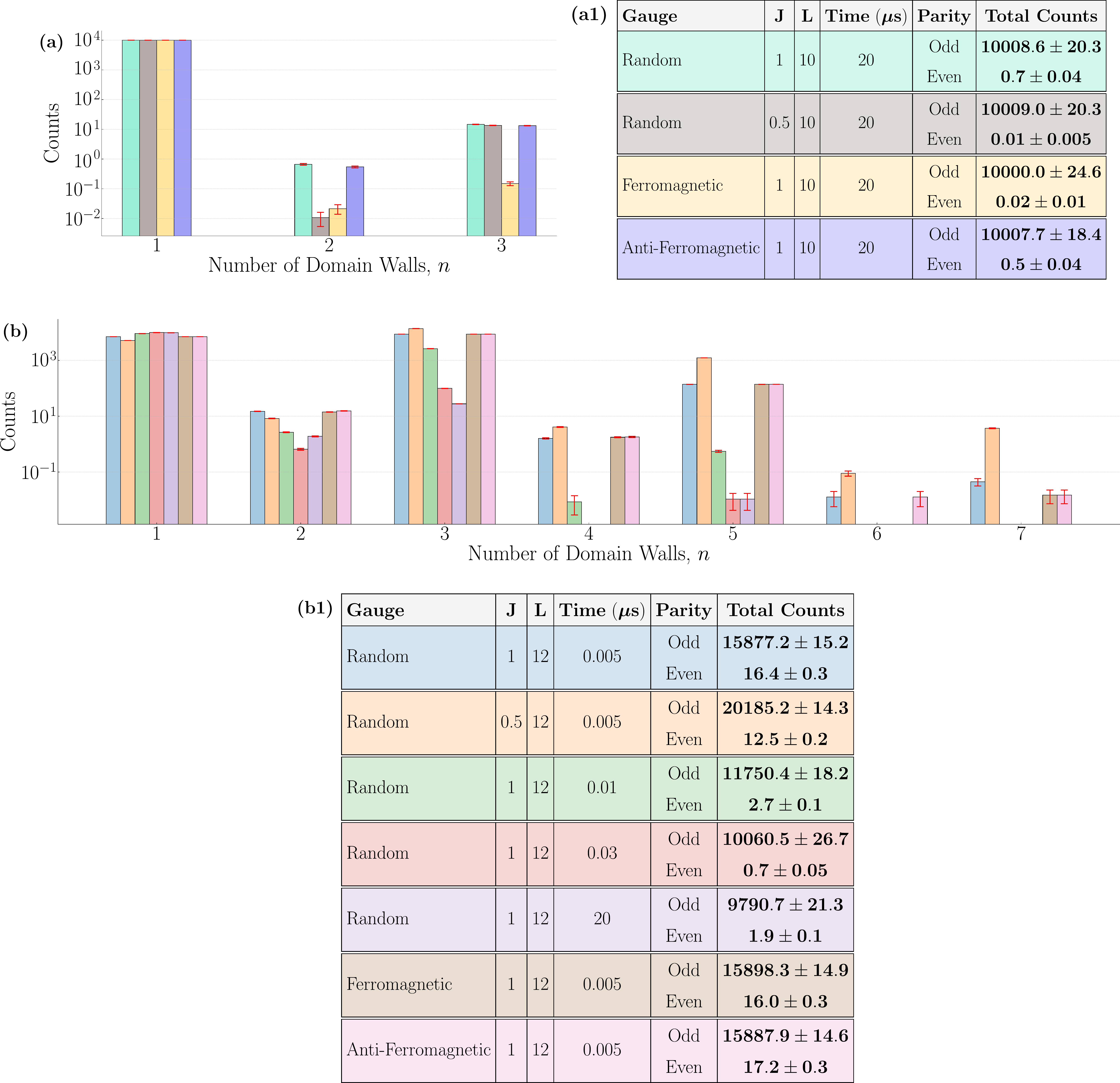}
\caption{\label{fig:domain_wall_plot_parity} 
Occurrence of multiple domain walls for each experimental setting, 
averaged over $468 \times 4$ chains for standard anneal protocol (a) and $562 \times 4$ chains for fast anneal protocol (b) of data with $10\,000$, is shown by separating configurations into those containing an even or odd number of domain walls for chains with antiparallel boundary conditions. 
The histograms show the total number of domain wall counts summed across the domain wall positions with the error bar representing combined uncertainty contributions in quadrature: the propagated binomial uncertainty on the mean across chains due to the finite number of domain wall occurrences, and the standard error of the mean arising from variations between different chains (for details see). The table reports the total number of domain walls obtained by summing the histogram counts, with the odd and even contributions combined separately and their uncertainties propagated in quadrature. The histogram and table in panels \textbf{(a)} and \textbf{(a1)} show the data for the standard anneal protocol, while the lower panel containing \textbf{(b)} and \textbf{(b1)} shows the corresponding results for the fast anneal protocol.}
\end{figure*}

In Fig.~\ref{fig:domain_multi_wall}, we present probability distribution from runs on the QPU for a chain length of $10$ in the standard anneal protocol for $20\,\mu$s and chain length of $12$ in the fast anneal protocol for $0.005\,\mu$s to compare between single and all domain wall occurrences. With antiparallel boundary conditions, the ground-state manifold consists of configurations containing a single domain wall, while higher odd numbers of domain walls correspond to excited states. Annealing at very fast anneal times gives the system very little time to settle into the lowest energy manifold with the system exciting into higher energy states. In rare cases (see table~\ref{fig:domain_wall_plot_parity}), even domain wall occurrences are also seen. For the fastest anneal time at $0.005\,\mu$s, a total of $15893.6 \pm 17.8$ domain wall occurrences are reported averaged across $4 \times 468$ chains in $10\,000$ experimental repeats on the QPU. Only $7072.9 \pm 11.2$ out of these had a single domain wall occurrence. Similarly, for $0.01\,\mu$s, we report $11753.1 \pm 19.1$ total domain wall occurrences with $9119.4 \pm 17.4$ of them being single domain wall occurrence. The numbers therefore support the  interpretation that during a fast quench the system undergoes diabatic evolution giving less time for defects to annihilate causing multiple domain walls survive to readout as the system no longer samples from the nearly degenerate single domain wall subspace. As we go higher up in the anneal times, $0.03\,\mu$s and $20\,\mu$s, the system transitions into mostly thermalized behavior. Filtering out configurations with multiple domain walls at these times scales makes little difference to the observed counts, as most configurations have single domain wall occurrences. table~\ref{tab:domain_count} contains the occurrences of the domain wall under different experimental settings.

\onecolumngrid
\section{Experimental data}\label{app:data}

Tables giving the averaged counts of domain walls in experimental and simulation runs, and related data.
The full data archive will be made available after publication \cite{OurRawData}.

\begin{table*}[tb!]
\centering
\renewcommand{\arraystretch}{2}

\newcommand{\paramcell}[3]{%
\begin{tabular}{c}
$#1~\mu$s \\ \hline
$J = #2$ \\ \hline
$#3$
\end{tabular}%
}

\resizebox{\textwidth}{!}{%
\begin{tabular}{|c|c|*{7}{c|}}
\hline
\multirow{2}{*}{\makecell{Domain wall \\ position, $i$}}
& \multirow{2}{*}{Data}
& \multicolumn{1}{|>{\columncolor{SetA!18}}c|}{\paramcell{0.005}{1}{\text{Random}}}
& \multicolumn{1}{>{\columncolor{SetB!18}}c|}{\paramcell{0.005}{0.5}{\text{Random}}}
& \multicolumn{1}{>{\columncolor{SetC!18}}c|}{\paramcell{0.01}{1}{\text{Random}}}
& \multicolumn{1}{>{\columncolor{SetD!18}}c|}{\paramcell{0.03}{1}{\text{Random}}}
& \multicolumn{1}{>{\columncolor{SetE!18}}c|}{\paramcell{20}{1}{\text{Random}}}
& \multicolumn{1}{>{\columncolor{SetF!18}}c|}{\paramcell{0.005}{1}{\text{Ferromagnetic}}}
& \multicolumn{1}{>{\columncolor{SetG!18}}c|}{\paramcell{0.005}{1}{\text{Anti-Ferromagnetic}}}

\\
\hline
\hline

\multirow{2}{*}{$1$}
& Raw      & $965.1 \pm 5.8$ & $1485.8 \pm 6.5$ & $673.5 \pm 5.3$ & $1029.6 \pm 9.8$ & $1010.9 \pm 8.4$ & $947.1 \pm 5.6$ & $935.5 \pm 5.5$ \\
\cline{2-9}
& Filtered & $370.3 \pm 2.7$ & $334.2 \pm 2.0$ & $470.2 \pm 4.0$ & $1017.3 \pm 9.7$ & $1008.1 \pm 8.4$ & $355.0 \pm 2.5$ & $356.2 \pm 2.5$ \\
\hline
\hline
\multirow{2}{*}{$2$}
& Raw      & $1710.9 \pm 6.2$ & $2173.5 \pm 5.9$ & $1000.2 \pm 6.2$ & $964.8 \pm 8.5$ & $947.6 \pm 7.2$ & $1699.1 \pm 6.0$ & $1714.2 \pm 5.8$ \\
\cline{2-9}
& Filtered & $743.0 \pm 3.5$ & $570.1 \pm 2.5$ & $742.2 \pm 5.0$ & $955.3 \pm 8.4$ & $944.8 \pm 7.2$ & $728.7 \pm 3.3$ & $741.4 \pm 3.3$ \\
\hline
\hline
\multirow{2}{*}{$3$}
& Raw      & $1679.7 \pm 5.8$ & $1931.0 \pm 4.6$ & $1147.9 \pm 6.2$ & $905.1 \pm 7.8$ & $875.2 \pm 5.9$ & $1682.0 \pm 5.6$ & $1700.0 \pm 5.7$ \\
\cline{2-9}
& Filtered & $803.2 \pm 3.7$ & $524.5 \pm 2.4$ & $899.4 \pm 5.5$ & $897.1 \pm 7.8$ & $872.6 \pm 5.9$ & $799.5 \pm 3.6$ & $810.2 \pm 3.7$ \\
\hline
\hline
\multirow{2}{*}{$4$}
& Raw      & $1500.2 \pm 4.9$ & $1820.8 \pm 3.6$ & $1226.5 \pm 5.9$ & $849.2 \pm 7.1$ & $838.8 \pm 5.3$ & $1506.1 \pm 5.0$ & $1509.5 \pm 4.8$ \\
\cline{2-9}
& Filtered & $710.2 \pm 3.6$ & $469.4 \pm 2.3$ & $986.2 \pm 5.8$ & $841.6 \pm 7.1$ & $836.2 \pm 5.3$ & $712.8 \pm 3.6$ & $714.0 \pm 3.4$ \\
\hline
\hline
\multirow{2}{*}{$5$}
& Raw      & $1422.5 \pm 4.3$ & $1794.9 \pm 3.3$ & $1212.0 \pm 5.2$ & $841.4 \pm 7.3$ & $814.5 \pm 5.1$ & $1424.5 \pm 4.1$ & $1399.0 \pm 4.1$ \\
\cline{2-9}
& Filtered & $630.4 \pm 3.4$ & $449.5 \pm 2.3$ & $970.5 \pm 5.5$ & $833.0 \pm 7.3$ & $812.0 \pm 5.1$ & $633.5 \pm 3.3$ & $615.0 \pm 3.1$ \\
\hline
\hline
\multirow{2}{*}{$6$}
& Raw      & $1378.6 \pm 4.1$ & $1789.1 \pm 3.4$ & $1217.8 \pm 5.2$ & $848.6 \pm 7.2$ & $811.1 \pm 5.0$ & $1385.5 \pm 4.3$ & $1369.6 \pm 4.2$ \\
\cline{2-9}
& Filtered & $589.8 \pm 3.1$ & $446.5 \pm 2.3$ & $973.1 \pm 5.7$ & $839.7 \pm 7.2$ & $808.4 \pm 5.0$ & $594.2 \pm 3.3$ & $584.1 \pm 3.1$ \\
\hline
\hline
\multirow{2}{*}{$7$}
& Raw      & $1383.6 \pm 4.3$ & $1789.8 \pm 3.3$ & $1221.2 \pm 5.2$ & $854.6 \pm 7.3$ & $822.3 \pm 5.2$ & $1399.1 \pm 4.1$ & $1398.9 \pm 4.1$ \\
\cline{2-9}
& Filtered & $609.1 \pm 3.3$ & $452.3 \pm 2.3$ & $979.2 \pm 5.5$ & $846.5 \pm 7.4$ & $819.7 \pm 5.2$ & $614.8 \pm 3.3$ & $610.5 \pm 3.1$ \\
\hline
\hline
\multirow{2}{*}{$8$}
& Raw      & $1502.4 \pm 4.8$ & $1813.0 \pm 3.5$ & $1207.6 \pm 5.8$ & $862.0 \pm 7.2$ & $850.3 \pm 5.6$ & $1514.3 \pm 4.9$ & $1509.7 \pm 4.8$ \\
\cline{2-9}
& Filtered & $709.3 \pm 3.5$ & $463.6 \pm 2.3$ & $968.4 \pm 5.6$ & $854.4 \pm 7.2$ & $847.5 \pm 5.6$ & $717.0 \pm 3.6$ & $718.0 \pm 3.5$ \\
\hline
\hline
\multirow{2}{*}{$9$}
& Raw      & $1715.9 \pm 6.1$ & $1936.9 \pm 4.6$ & $1137.2 \pm 6.0$ & $907.7 \pm 7.8$ & $881.7 \pm 6.1$ & $1732.9 \pm 6.0$ & $1678.7 \pm 5.5$ \\
\cline{2-9}
& Filtered & $818.0 \pm 3.9$ & $528.7 \pm 2.4$ & $890.2 \pm 5.3$ & $899.7 \pm 7.8$ & $879.0 \pm 6.1$ & $826.5 \pm 3.8$ & $799.3 \pm 3.5$ \\
\hline
\hline
\multirow{2}{*}{$10$}
& Raw      & $1687.6 \pm 6.4$ & $2179.5 \pm 6.1$ & $1021.3 \pm 6.5$ & $973.4 \pm 8.7$ & $936.4 \pm 7.1$ & $1698.5 \pm 6.1$ & $1734.0 \pm 6.3$ \\
\cline{2-9}
& Filtered & $729.3 \pm 3.6$ & $572.1 \pm 2.6$ & $759.8 \pm 5.2$ & $963.4 \pm 8.6$ & $933.7 \pm 7.1$ & $732.3 \pm 3.3$ & $753.0 \pm 3.5$ \\
\hline
\hline
\multirow{2}{*}{$11$}
& Raw      & $947.1 \pm 5.7$ & $1483.3 \pm 6.3$ & $688.0 \pm 5.7$ & $1024.8 \pm 9.6$ & $1003.7 \pm 8.6$ & $925.1 \pm 5.5$ & $955.9 \pm 5.5$ \\
\cline{2-9}
& Filtered & $360.2 \pm 2.6$ & $334.4 \pm 1.9$ & $480.3 \pm 4.2$ & $1012.7 \pm 9.5$ & $1000.9 \pm 8.5$ & $348.6 \pm 2.4$ & $364.9 \pm 2.5$ \\
\hline
\hline
\multirow{2}{*}{$\mathrm{Total}$}
& Raw      & $\mathbf{15893.6 \pm 17.8}$ & $\mathbf{20197.6 \pm 16.0}$ & $\mathbf{11753.1 \pm 19.1}$ & $\mathbf{10061.1 \pm 26.8}$ & $\mathbf{9792.6 \pm 21.3}$ & $\mathbf{15914.3 \pm 17.4}$ & $\mathbf{15905.2 \pm 17.1}$ \\
\cline{2-9}
& Filtered & $\mathbf{7072.9 \pm 11.2}$ & $\mathbf{5145.4 \pm 7.6}$ & $\mathbf{9119.4 \pm 17.4}$ & $\mathbf{9960.9 \pm 26.7}$ & $\mathbf{9762.9 \pm 21.3}$ & $\mathbf{7062.9 \pm 10.9}$ & $\mathbf{7066.7 \pm 10.7}$ \\ 
\hline

\end{tabular}%
}

\caption{The table contains the total number of domain wall counts at each domain wall position and the total, summed over all positions, across $4$ batches ($4 \times 468$ chains) of data, with chain length $L=12$, with $10\,000$  experimental repetitions for the fast anneal protocol for different annealing parameters (gauge, coupling strength $J$ and anneal time) on a $12$ spin Ising chain.  We report both the raw data containing possible single and multiple domain walls as it is possible for chains to have a multiple domain walls at different positions, and the filtered data containing only single domain wall where only chains with a single domain wall are kept and the others discarded. We also report the uncertainty associated with each value. The uncertainty primarily combines two contributions in quadrature: the propagated binomial uncertainty on the mean across chains due to the finite number of domain wall occurrences within each batch, and the standard error of the mean arising from variations between different chains in each batch. Then as mean values from the four batches are added together, the combined uncertainties from the four batches are added in quadrature again }
\label{tab:domain_count}
\end{table*}

\begin{table*}[tb!]
\centering
\renewcommand{\arraystretch}{2}

\newcommand{\paramcell}[3]{%
\begin{tabular}{c}
$#1~\mu$s \\ \hline
$J = #2$ \\ \hline
$#3$
\end{tabular}%
}

\resizebox{\textwidth}{!}{%
\setlength{\tabcolsep}{12pt}
\begin{tabular}{|c|c|*{4}{c|}}
\hline
\multirow{2}{*}{\makecell{Domain wall \\ position, $i$}}
& \multirow{2}{*}{Data}
& \multicolumn{1}{|>{\columncolor{SetA1!18}}c|}{\paramcell{20}{1}{\text{Random}}}
& \multicolumn{1}{>{\columncolor{SetB1!18}}c|}{\paramcell{20}{0.5}{\text{Random}}}
& \multicolumn{1}{>{\columncolor{SetC1!18}}c|}{\paramcell{20}{1}{\text{Ferromagnetic}}}
& \multicolumn{1}{>{\columncolor{SetD1!18}}c|}{\paramcell{20}{1}{\text{Anti-Ferromagnetic}}}

\\
\hline
\hline

\hline
\multirow{2}{*}{$1$}
& Raw      & $1168.0 \pm 8.6$ & $992.8 \pm 7.5$ & $823.4 \pm 7.9$ & $1150.1 \pm 7.7$ \\
\cline{2-6}
& Filtered & $1166.4 \pm 8.6$ & $991.6 \pm 7.5$ & $823.4 \pm 7.9$ & $1148.7 \pm 7.7$ \\
\hline
\multirow{2}{*}{$2$}
& Raw      & $1166.7 \pm 7.1$ & $1231.3 \pm 7.8$ & $1177.5 \pm 8.9$ & $1144.3 \pm 6.4$ \\
\cline{2-6}
& Filtered & $1165.0 \pm 7.1$ & $1229.6 \pm 7.8$ & $1177.5 \pm 8.9$ & $1142.8 \pm 6.4$ \\
\hline
\multirow{2}{*}{$3$}
& Raw      & $1080.7 \pm 5.9$ & $1146.2 \pm 6.4$ & $1163.5 \pm 7.9$ & $1089.0 \pm 5.6$ \\
\cline{2-6}
& Filtered & $1079.0 \pm 5.9$ & $1144.6 \pm 6.4$ & $1163.5 \pm 7.9$ & $1087.5 \pm 5.6$ \\
\hline
\multirow{2}{*}{$4$}
& Raw      & $1057.2 \pm 5.5$ & $1110.3 \pm 5.6$ & $1107.7 \pm 7.4$ & $1046.7 \pm 4.9$ \\
\cline{2-6}
& Filtered & $1055.5 \pm 5.5$ & $1108.7 \pm 5.6$ & $1107.7 \pm 7.4$ & $1045.1 \pm 4.9$ \\
\hline
\multirow{2}{*}{$5$}
& Raw      & $1049.8 \pm 5.5$ & $1084.3 \pm 5.6$ & $1127.5 \pm 6.9$ & $1043.9 \pm 4.7$ \\
\cline{2-6}
& Filtered & $1048.1 \pm 5.5$ & $1082.8 \pm 5.6$ & $1127.4 \pm 6.9$ & $1042.3 \pm 4.7$ \\
\hline
\multirow{2}{*}{$6$}
& Raw      & $1056.3 \pm 5.4$ & $1103.3 \pm 5.7$ & $1133.3 \pm 7.3$ & $1058.2 \pm 4.9$ \\
\cline{2-6}
& Filtered & $1054.6 \pm 5.4$ & $1101.7 \pm 5.7$ & $1133.2 \pm 7.3$ & $1056.7 \pm 4.9$ \\
\hline
\multirow{2}{*}{$7$}
& Raw      & $1092.9 \pm 6.2$ & $1145.3 \pm 6.4$ & $1221.3 \pm 8.3$ & $1109.7 \pm 5.6$ \\
\cline{2-6}
& Filtered & $1091.1 \pm 6.2$ & $1143.8 \pm 6.4$ & $1221.3 \pm 8.3$ & $1108.2 \pm 5.6$ \\
\hline
\multirow{2}{*}{$8$}
& Raw      & $1163.0 \pm 7.1$ & $1206.5 \pm 7.7$ & $1304.8 \pm 10.1$ & $1172.0 \pm 6.6$ \\
\cline{2-6}
& Filtered & $1161.3 \pm 7.1$ & $1204.9 \pm 7.7$ & $1304.8 \pm 10.1$ & $1170.4 \pm 6.6$ \\
\hline
\multirow{2}{*}{$9$}
& Raw      & $1174.7 \pm 8.5$ & $989.0 \pm 7.7$ & $941.1 \pm 8.7$ & $1194.2 \pm 7.9$ \\
\cline{2-6}
& Filtered & $1173.1 \pm 8.5$ & $987.7 \pm 7.7$ & $941.1 \pm 8.7$ & $1192.7 \pm 7.9$ \\
\hline
\multirow{2}{*}{$\mathrm{Total}$}
& Raw      & $\mathbf{10009.3 \pm 20.3}$ & $\mathbf{10009.0 \pm 20.3}$ & $\mathbf{10000.0 \pm 24.6}$ & $\mathbf{10008.2 \pm 18.4}$ \\
\cline{2-6}
& Filtered & $\mathbf{9994.0 \pm 20.3}$ & $\mathbf{9995.5 \pm 20.3}$ & $\mathbf{9999.8 \pm 24.6}$ & $\mathbf{9994.4 \pm 18.4}$ \\

\hline
\end{tabular}%
}
\caption{The table contains the total number of domain wall counts at each domain wall position and the total, summed over all positions, across $4$ batches ($4 \times 562$ chains) of data, with chain length $L=10$, with $10\,000$ experimental repetitions for the standard anneal protocol for different annealing parameters (gauge, coupling strength $J$ and anneal time) on a $10$ spin Ising chain.  We report both the raw data containing possible single and multiple domain walls as it is possible for chains to have a multiple domain walls at different positions, and the filtered data containing only single domain wall where only chains with a single domain wall are kept and the others discarded. We also report the uncertainty associated with each value. The uncertainty primarily combines two contributions in quadrature: the propagated binomial uncertainty on the mean across chains due to the finite number of domain wall occurrences within each batch, and the standard error of the mean arising from variations between different chains in each batch. Then as mean values from the four batches are added together, the combined uncertainties from the four batches are added in quadrature again.}
\label{tab:domain_count1}
\end{table*}


\begin{table}[tb!]
\centering
\renewcommand{\arraystretch}{1.35}
\setlength{\tabcolsep}{8pt}
\definecolor{SetAA}{HTML}{39FF14} 
\definecolor{SetBB}{HTML}{C7FF00}

\begin{tabular}{llcccc}
\hline
\renewcommand{\arraystretch}{1.4}%
\textbf{Dataset} 
&
& \textbf{Left } 
& \textbf{Right} 
& \textbf{Either} 
& \textbf{Both} \\
\hline

\cellcolor{SetA1!18}\multirow{2}{*}
& Raw 
& $0.56 \pm 0.05$ & $0.57 \pm 0.05$ & $1.13 \pm 0.07$ & $0.00 \pm 0.00$  \\
\cline{2-6}
\cellcolor{SetA1!18}\raisebox{1.75ex}{$20~\mu$s, Random}
& Filtered
& $0.00 \pm 0.00$ & $0.00 \pm 0.00$ & $0.00 \pm 0.00$ & $0.00 \pm 0.00$  \\
\hline
\hline

\cellcolor{SetB1!18}\multirow{2}{*}
& Raw
& $0.007 \pm 0.005$ & $0.007 \pm 0.005$ & $0.014 \pm 0.007$ & $0.0000 \pm 0.0000$ \\
\cline{2-6}
\cellcolor{SetB1!18}\raisebox{1.75ex}{$20~\mu$s, $J=0.5$, Random}
& Filtered
& $0.00 \pm 0.00$ & $0.00 \pm 0.00$ & $0.00 \pm 0.00$ & $0.0000 \pm 0.00$  \\
\hline
\hline

\cellcolor{SetC1!18}\multirow{2}{*}
& Raw
& $0.07 \pm 0.02$ & $0.07 \pm 0.02$ & $0.14 \pm 0.02$ & $0.00 \pm 0.00$ \\

\cline{2-6}

\cellcolor{SetC1!18}\raisebox{1.75ex}{$20~\mu$s, Ferromagnetic}
& Filtered
& $0.00 \pm 0.00$
& $0.00 \pm 0.00$
& $0.00 \pm 0.00$
& $0.00 \pm 0.00$ \\
\hline
\hline

\cellcolor{SetD1!18}\multirow{2}{*}
& Raw
& $0.56 \pm 0.05$ & $0.64 \pm 0.05$ & $1.20 \pm 0.07$ & $0.00 \pm 0.00$ \\

\cline{2-6}
\cellcolor{SetD1!18}\raisebox{1.75ex}{$20~\mu$s, Antiferromagnetic}
& Filtered
& $0.00 \pm 0.00$
& $0.00 \pm 0.00$
& $0.00 \pm 0.00$
& $0.00 \pm 0.00$ \\
\hline

\end{tabular}

\vspace{4mm}

\begin{tabular}{llcccc}
\hline
\renewcommand{\arraystretch}{1.6}%

\textbf{Dataset} 
&
& \textbf{Left} 
& \textbf{Right} 
& \textbf{Either} 
& \textbf{Both} \\
\hline
\cellcolor{SetA!19}\multirow{2}{*}
& Raw
& $5.2 \pm 0.2$ & $5.6 \pm 0.3$ & $10.8 \pm 0.4$ & $0.009 \pm 0.006$  \\
\cline{2-6}

\cellcolor{SetA!19}\raisebox{1.75ex}{$0.005~\mu$s, Random}
& Filtered
& $0.009 \pm 0.006$ & $0.009 \pm 0.006$ & $0.009 \pm 0.006$ & $0.009 \pm 0.006$  \\
\hline
\hline
\cellcolor{SetB!19}\multirow{2}{*}

& Raw & $3.1 \pm 0.3$ & $2.9 \pm 0.2$ & $6.0 \pm 0.4$ & $0.00 \pm 0.00$ \\
\cline{2-6}
\cellcolor{SetB!19}\raisebox{1.75ex}{$0.005~\mu$s, $J=0.5$, Random}
& Filtered
& $0.00 \pm 0.00$ & $0.00 \pm 0.00$ & $0.00 \pm 0.00$ & $0.00 \pm 0.00$ \\
\hline
\hline
\cellcolor{SetC!19}\multirow{2}{*}

& Raw & $1.8 \pm 0.1$ & $1.8 \pm 0.1$ & $3.7 \pm 0.1$ & $0.00 \pm 0.00$  \\
\cline{2-6}
\cellcolor{SetC!19}\raisebox{1.75ex}{$0.01~\mu$s, Random}
& Filtered
& $0.00 \pm 0.00$ & $0.00 \pm 0.00$ & $0.00 \pm 0.00$ & $0.00 \pm 0.00$ \\
\hline
\hline
\cellcolor{SetD!19}\multirow{2}{*}

& Raw & $3.1 \pm 0.1$ & $2.8 \pm 0.1$ & $5.9 \pm 0.2$ & $0.00 \pm 0.00$ \\
\cline{2-6}
\cellcolor{SetD!19}\raisebox{1.75ex}{$0.03~\mu$s, Random}
& Filtered
& $0.00 \pm 0.00$ & $0.00 \pm 0.00$ & $0.00 \pm 0.00$ & $0.00 \pm 0.00$ \\

\hline
\hline
\cellcolor{SetE!19}\multirow{2}{*}

& Raw & $116 \pm 2$ & $112 \pm 2$ & $228 \pm 3$ & $0.02 \pm 0.01$ \\
\cline{2-6}
\cellcolor{SetE!19}\raisebox{1.75ex}{$20~\mu$s, Random}
& Filtered
& $0.02 \pm 0.01$ & $0.02 \pm 0.01$ & $0.02 \pm 0.01$ & $0.02 \pm 0.01$  \\
\hline
\hline

\cellcolor{SetF!19}\multirow{2}{*}
& Raw & $5.3 \pm 0.3$ & $5.2 \pm 0.3$ & $10.5 \pm 0.5$ & $0.006 \pm 0.005$   \\
\cline{2-6}
\cellcolor{SetF!19}\raisebox{1.75ex}{$0.005~\mu$s, Ferromagnetic}
& Filtered
& $0.004 \pm 0.004$ & $0.004 \pm 0.004$ & $0.004 \pm 0.004$ & $0.004 \pm 0.004$ \\ 
\hline
\hline

\cellcolor{SetG!19}\multirow{2}{*}
& Raw & $5.7 \pm 0.3$ & $5.7 \pm 0.3$ & $11.5 \pm 0.5$ & $0.002 \pm 0.003$   \\
\cline{2-6}
\cellcolor{SetG!19}\raisebox{1.75ex}{$0.005~\mu$s, Antiferromagnetic}
& Filtered
& $0.00 \pm 0.00$ & $0.00 \pm 0.00$ & $0.00 \pm 0.00$ & $0.00 \pm 0.00$ \\

\hline

\end{tabular}
\caption{Boundary spin failure counts out of $10\,000$ experimental repetitions per chain for each experimental setting, averaged over $562$ chains with chain length $L=10$ for standard anneal protocol (top panel) and $468$ chains with chain length $L=10$ for fast anneal protocol (bottom panel).The table includes four cases: spin flip (pinned boundary failure) counts in the left boundary, in the right boundary, at either the left or the right boundary, and at both ends. Either boundary counts shots where at least one boundary spin has flipped and both boundaries counts only occurrences where both boundary spins have flipped. Each entry has total uncertainty associated with it and is calculated by combining two sources of uncertainties, binomial error arising from finite reads and standard error accounting for variation among the chains, in quadrature.}
\label{tab:flip}
\vspace{2mm}
\end{table}

\end{document}